\documentclass[prd,superscriptaddress,unsortedaddress,twocolumn,showpacs,preprintnumbers,amsmath,amssymb]{revtex4}
\usepackage[dvips]{graphicx}
\usepackage{amsmath,amssymb,times}

\newcommand{\bequ}{\begin{equation}}
\newcommand{\eequ}{\end{equation}}
\newcommand{\bea}{\begin{eqnarray}}
\newcommand{\eea}{\end{eqnarray}}

\DeclareSymbolFont{boldletters}{OML}{cmm} {b}{it}
\DeclareSymbolFontAlphabet{\mathbit}{boldletters}
\DeclareMathSymbol{\alpha}{\mathalpha}{letters}{"0B}
\DeclareMathSymbol{\beta}{\mathalpha}{letters}{"0C}
\DeclareMathSymbol{\gamma}{\mathalpha}{letters}{"0D}
\DeclareMathSymbol{\delta}{\mathalpha}{letters}{"0E}
\DeclareMathSymbol{\epsilon}{\mathalpha}{letters}{"0F}
\DeclareMathSymbol{\zeta}{\mathalpha}{letters}{"10}
\DeclareMathSymbol{\eta}{\mathalpha}{letters}{"11}
\DeclareMathSymbol{\theta}{\mathalpha}{letters}{"12}
\DeclareMathSymbol{\iota}{\mathalpha}{letters}{"13}
\DeclareMathSymbol{\kappa}{\mathalpha}{letters}{"14}
\DeclareMathSymbol{\lambda}{\mathalpha}{letters}{"15}
\DeclareMathSymbol{\mu}{\mathalpha}{letters}{"16}
\DeclareMathSymbol{\nu}{\mathalpha}{letters}{"17}
\DeclareMathSymbol{\xi}{\mathalpha}{letters}{"18}
\DeclareMathSymbol{\pi}{\mathalpha}{letters}{"19}
\DeclareMathSymbol{\rho}{\mathalpha}{letters}{"1A}
\DeclareMathSymbol{\sigma}{\mathalpha}{letters}{"1B}
\DeclareMathSymbol{\tau}{\mathalpha}{letters}{"1C}
\DeclareMathSymbol{\upsilon}{\mathalpha}{letters}{"1D}
\DeclareMathSymbol{\phi}{\mathalpha}{letters}{"1E}
\DeclareMathSymbol{\chi}{\mathalpha}{letters}{"1F}
\DeclareMathSymbol{\psi}{\mathalpha}{letters}{"20}
\DeclareMathSymbol{\omega}{\mathalpha}{letters}{"21}
\DeclareMathSymbol{\varepsilon}{\mathalpha}{letters}{"22}
\DeclareMathSymbol{\vartheta}{\mathalpha}{letters}{"23}
\DeclareMathSymbol{\varpi}{\mathalpha}{letters}{"24}
\DeclareMathSymbol{\varrho}{\mathalpha}{letters}{"25}
\DeclareMathSymbol{\varsigma}{\mathalpha}{letters}{"26}
\DeclareMathSymbol{\varphi}{\mathalpha}{letters}{"27}
\DeclareMathSymbol{\Gamma}{\mathalpha}{letters}{"00}
\DeclareMathSymbol{\Delta}{\mathalpha}{letters}{"01}
\DeclareMathSymbol{\Theta}{\mathalpha}{letters}{"02}
\DeclareMathSymbol{\Lambda}{\mathalpha}{letters}{"03}
\DeclareMathSymbol{\Xi}{\mathalpha}{letters}{"04}
\DeclareMathSymbol{\Pi}{\mathalpha}{letters}{"05}
\DeclareMathSymbol{\Sigma}{\mathalpha}{letters}{"06}
\DeclareMathSymbol{\Upsilon}{\mathalpha}{letters}{"07}
\DeclareMathSymbol{\Phi}{\mathalpha}{letters}{"08}
\DeclareMathSymbol{\Psi}{\mathalpha}{letters}{"09}
\DeclareMathSymbol{\Omega}{\mathalpha}{letters}{"0A}

\begin{document}
\preprint{SAGA-HE-248-08}
\title{Determination of QCD phase diagram from the imaginary 
chemical potential region}

\author{Yuji Sakai}
\email[]{sakai@phys.kyushu-u.ac.jp}
\affiliation{Department of Physics, Graduate School of Sciences, Kyushu University,
             Fukuoka 812-8581, Japan}
\author{Kouji Kashiwa}
\email[]{kashiwa@phys.kyushu-u.ac.jp}
\affiliation{Department of Physics, Graduate School of Sciences, Kyushu University,
             Fukuoka 812-8581, Japan}

\author{Hiroaki Kouno}
\email[]{kounoh@cc.saga-u.ac.jp}
\affiliation{Department of Physics, Saga University,
             Saga 840-8502, Japan}

\author{Masayuki Matsuzaki}
\email[]{matsuza@fukuoka-edu.ac.jp}
\affiliation{Department of Physics, Fukuoka University of Education, 
             Munakata, Fukuoka 811-4192, Japan}

\author{Masanobu Yahiro}
\email[]{yahiro@phys.kyushu-u.ac.jp}
\affiliation{Department of Physics, Graduate School of Sciences, Kyushu University,
             Fukuoka 812-8581, Japan}

\date{\today}

\begin{abstract}
We test the reliability of 
the the Polyakov-loop extended Nambu--Jona-Lasinio (PNJL) 
model, comparing the model result with the lattice data 
at nonzero imaginary chemical potential. 
The PNJL model with the vector-type four-quark and scalar-type eight-quark 
interactions reproduces the lattice data on 
the pseudocritical temperatures 
of the deconfinement and chiral phase transitions. 
The QCD phase diagram in the real chemical potential region 
is predicted by the PNJL model. The critical endpoint survives, even if 
the vector-type four-quark interaction is taken into account.

\end{abstract}

\maketitle

\section{Introduction}
\label{Introduction}

Quantum Chromodynamics (QCD) is a remarkable theory. It is
renormalizable and essentially parameter free.
QCD accounts for the rich phenomenology of hadronic and nuclear physics.
Thermodynamics of QCD is also well defined. Nevertheless, it is 
not well known because of its nonperturbative nature. 
In particular, QCD phase diagram is essential for 
understanding not only natural 
phenomena such as compact stars and the early universe but also 
laboratory experiments such as relativistic heavy-ion collisions. 

Unfortunately, quantitative calculations of the phase diagram from 
first-principle lattice QCD (LQCD) have the well known sign problem 
when the chemical potential ($\mu$) is real; 
for example, see Ref.~\cite{Kogut} and references therein. 
So far, several approaches have been proposed to circumvent the difficulty; 
for example, the reweighting method~\cite{Fodor}, 
the Taylor expansion method~\cite{Allton} and 
the analytic continuation to real chemical potential ($\mu_\mathrm{R}$) 
from imaginary chemical potential ($\mu_\mathrm{I}$)~
\cite{FP,FP3,Elia,Chen34,Chen}. 
However, those are still far from perfection.  

As an approach complementary to first-principle lattice QCD, 
we can consider effective models such as the  
Nambu--Jona-Lasinio (NJL) model~
\cite{NJ1,AY,BR,Sca,Fuj,KKKN,Osipov,Kashiwa,Sakaetal} and 
the Polyakov-loop extended Nambu--Jona-Lasinio (PNJL) 
model~\cite{Meisinger,Dumitru,Fukushima,Ghos,Megias,Ratti,Ciminale,
Rossner,Hansen,Sasaki,Schaefer,Costa,Kashiwa1,Fu,Abuki,Sakai,Fukushima2,
Sakai1}. 
The NJL model describes the chiral symmetry breaking, but not 
the confinement mechanism. 
The PNJL model 
is designed \cite{Fukushima} to make it possible to treat 
the Polyakov loop as well as the chiral symmetry breaking. 

In the NJL-type models, 
the input parameters are determined at $\mu=0$ and $T \ge 0$, 
where $T$ is temperature. 
It is then highly nontrivial whether the models predict properly dynamics 
of QCD at finite $\mu_\mathrm{R}$.  
This should be tested from QCD. 
Fortunately, this is possible in the $\mu_\mathrm{I}$ region, 
since lattice QCD has no sign problem there.
The canonical partition function $Z_{\rm C}(n)$ 
with real quark number $n$ is the Fourier transform of 
the grand-canonical one $Z_{\rm GC}(\theta)$ 
with $\theta=\mu_\mathrm{I}/T$~\cite{RW}: 
\bea
Z_{\rm C}(n)=\frac{1}{2\pi}\int_{-\pi}^{\pi} d\theta e^{-in\theta}
Z_{\rm GC}(\theta) .
\label{trans-to-n}
\eea 
Thus, the thermodynamic potential of QCD, 
$\Omega_{\rm QCD}(\theta)=-T \ln(Z_{\rm GC}(\theta))$, at finite $\theta$ 
includes all dynamics 
at real $n$ and hence at finite $\mu_\mathrm{R}$. 
Therefore, the reliability of effective models at finite $\mu_\mathrm{R}$ 
can be tested in the $\mu_\mathrm{I}$ region. 

Roberge and Weiss found~\cite{RW} that QCD has a periodicity 
$\Omega_{\rm QCD}(\theta)=\Omega_{\rm QCD}(\theta+2\pi k/3)$, 
showing that $\Omega_{\rm QCD}(\theta+2\pi k/3)$ is 
transformed into $\Omega_{\rm QCD}(\theta)$ by 
the ${\mathbb Z}_3$ transformation with integer $k$. 
This means that QCD is invariant under 
a combination of the ${\mathbb Z}_3$ transformation and 
a parameter transformation $\theta \to \theta+2k\pi/3$~\cite{Sakai,Sakai1}, 
\bea
q &\to& Uq, \quad 
A_{\nu} \to UA_{\nu}U^{-1} - i/g (\partial_{\nu}U)U^{-1} , \nonumber \\ 
\theta &\to& \theta+2\pi k/3 ,
\label{ez3}
\eea
where $U(x,\tau)$ are elements of SU(3) with 
$
U(x,\beta=1/T)=\exp(-2i \pi k/3)U(x,0)   
$
and $q$ is the quark field. 
We call this combination the extended 
${\mathbb Z}_3$ transformation. 
Thus, $\Omega_{\rm QCD}(\theta)$ 
has the extended ${\mathbb Z}_3$ symmetry, and hence 
quantities invariant under the extended ${\mathbb Z}_3$ transformation
have the RW periodicity~\cite{Sakai,Sakai1}. 
At the present stage, the PNJL model is 
only a realistic effective model that possesses both 
the extended ${\mathbb Z}_3$ symmetry and 
chiral symmetry~\cite{Sakai,Sakai1}. 
This property makes it possible to compare PNJL with lattice QCD 
quantitatively in the $\mu_\mathrm{I}$ region. 
If the PNJL model succeeds in reproducing the lattice data, we may think that 
the PNJL model will predict, with high reliability, the QCD phase structure 
in the $\mu_\mathrm{R}$ region. 

The extended ${\mathbb Z}_3$ symmetry in QCD 
is a remnant of the ${\mathbb Z}_3$ symmetry, namely 
the confinement mechanism, in the pure gauge system. 
The extended ${\mathbb Z}_3$ symmetry appears as the RW periodicity in 
the $\mu_\mathrm{I}$ region and implicitly affects dynamics 
in the $\mu_\mathrm{R}$ region.    
Actually, the mechanism largely shifts 
the critical endpoint~\cite{AY} toward  higher $T$ and lower $\mu$ 
than the NJL model predicts~\cite{Rossner,Kashiwa1,Fukushima2}. 
In contrast, the vector-type four-quark interaction 
$G_{\rm v}(\bar{q}\gamma_\mu q)^2$ 
largely moves the critical endpoint 
in the opposite direction~\cite{KKKN,Kashiwa,Kashiwa1,Fukushima2}, 
if it is newly added to the NJL and PNJL models. Thus, 
it is essential to determine the strength 
of the coupling $G_{\rm v}$ 
of the vector-type interaction, 
although the interaction is often ignored in the NJL and PNJL calculations. 

In the relativistic meson-nucleon theory~\cite{Walecka}, 
the repulsive force mediated by vector mesons 
is essential to account for the saturation property of nuclear matter. 
Using the auxiliary field method, one 
can convert quark-quark interactions 
to meson-quark interactions; 
for example, see Refs.~\cite{KS,Sakaetal,KSKHTMY} and references therein. 
In the hadron phase, quarks have a large effective mass as a result 
of spontaneous chiral symmetry breaking, and then nucleons can be 
considered to be formed by such three heavy quarks, i.e. 
three constituent quarks. 
It is then natural to think that 
there exists the correspondence between the meson-nucleon interactions and 
the quark-quark interactions. 
In this sense, it is very likely that the vector-type four-quark interaction 
is not negligible and even significant in particular 
at a finite quark-density region corresponding 
to the nuclear saturation density. 
In the previous work~\cite{Sakai1}, we have proposed 
that the strength of $G_{\rm v}$ can be determined from lattice data on 
the chiral phase transition in the $\mu_\mathrm{I}$ region. 

In this paper, we consider two-flavor QCD and 
show the reliability of the PNJL model, quantitatively comparing 
the model result with lattice data in the $\mu_\mathrm{I}$ region. 
The model parameters except $G_{\rm v}$ are fixed 
by the measured pion mass and decay constant at $\mu=T=0$ and 
lattice data~\cite{Karsch4,Karsch3,Kaczmarek2} at $T > 0$ and $\mu=0$. 
The PNJL calculation with no vector-type interaction well reproduces 
lattice data~\cite{FP,Chen} on the pseudocritical temperature $T_\mathrm{c}(\Phi)$ of 
the deconfinement phase transition, but not 
on the pseudocritical temperature $T_\mathrm{c}(\sigma)$ of 
the chiral phase transition near $\theta=\pi/3$. 
The strength of $G_{\rm v}$ is fitted so as to reproduce the latter data. 
The primary result of the lattice simulations is that 
$T_\mathrm{c}(\Phi)$ coincides with $T_\mathrm{c}(\sigma)$, within numerical errors, 
in the entire region of $\theta$~\cite{FP,Chen}. 
The PNJL model with the vector-type interaction can reproduce this property. 
Finally, we quantitatively predict the phase diagram in the $\mu_\mathrm{R}$ 
region by using the PNJL model with the parameter set justified 
in the $\mu_\mathrm{I}$ region. 
This sort of model predictions are quite important 
before doing heavy lattice calculations with large lattice size 
in the $\mu_\mathrm{I}$ region. 

In section II, the PNJL model is explained simply. 
In section III, we test the PNJL model in the $\mu_\mathrm{I}$ region and 
determine the strength of $G_{\rm v}$. Finally, we predict 
the phase diagram in the $\mu_\mathrm{R}$ region. 
Section IV is devoted to summary.

\section{PNJL model}
\label{PNJL}

The two-flavor PNJL Lagrangian is 
\begin{align}
 {\cal L}  =& {\bar q}(i \gamma_\nu D^\nu -m_0)q \notag\\
             &\hspace{3mm} + G_{\rm s}[({\bar q}q)^2 
                          +({\bar q}i\gamma_5 {\vec \tau}q)^2] 
              - {\cal U}(\Phi [A],{\Phi} [A]^*,T) ,
             \label{eq:E1}
\end{align}
where $q$ denotes the two-flavor quark field, 
$m_0$ does the current quark mass, and 
$D^\nu=\partial^\nu+iA^\nu-i\mu\delta^{\nu}_{0}$. 
The field $A^\nu$ is defined as 
$A^\nu=\delta^{\nu}_{0}gA^0_a{\lambda^a\over{2}}$
with the gauge field $A^\nu_a$, 
the Gell-Mann matrix $\lambda_a$ and the gauge coupling $g$.
In the NJL sector, 
$G_{\rm s}$ denotes the coupling constant of the scalar-type 
four-quark interaction. 
Later, we will add 
the vector-type four-quark interaction  \cite{AY,KKKN,Kashiwa,Sakai1} and 
the scalar-type eight-quark 
interaction~\cite{Osipov,Kashiwa,Sakai} to the PNJL Lagrangian. 
The Polyakov potential ${\cal U}$, defined in (\ref{eq:E13}), 
is a function of the Polyakov loop $\Phi$ and its Hermitian 
conjugate $\Phi^*$,
\begin{align}
\Phi      = {1\over{N_{\rm c}}}{\rm Tr} L,~~~~
\Phi^{*}  = {1\over{N_{\rm c}}} {\rm Tr}L^\dag ,
\end{align}
with
\begin{align}
L({\bf x}) = {\cal P} \exp\Bigl[
                {i\int^\beta_0 d \tau A_4({\bf x},\tau)}\Bigr],
\end{align}
where ${\cal P}$ is the path ordering and $A_4 = iA_0 $. 
In the chiral limit ($m_0=0$), 
the Lagrangian density has the exact 
$SU(N_f)_{\rm L} \times SU(N_f)_{\rm R}
\times U(1)_{\rm v} \times SU(3)_{\rm c}$  symmetry. 

The temporal component of the gauge field is diagonal 
in the flavor space, because the color and the flavor space 
are completely separated out in the present case. 
In the Polyakov gauge, $L$ can be written in a diagonal form 
in the color space~\cite{Fukushima}: 
\begin{align}
L 
=  e^{i \beta (\phi_3 \lambda_3 + \phi_8 \lambda_8)}
= {\rm diag} (e^{i \beta \phi_a},e^{i \beta \phi_b},
e^{i \beta \phi_c} ),
\label{eq:E6}
\end{align}
where $\phi_a=\phi_3+\phi_8/\sqrt{3}$, $\phi_b=-\phi_3+\phi_8/\sqrt{3}$
and $\phi_c=-(\phi_a+\phi_b)=-2\phi_8/\sqrt{3}$. 
The Polyakov loop $\Phi$ is an exact order parameter of the spontaneous 
${\mathbb Z}_3$ symmetry breaking in the pure gauge theory.
Although the ${\mathbb Z}_3$ symmetry is not an exact one 
in the system with dynamical quarks, it still seems to be a good indicator of 
the deconfinement phase transition. 
Therefore, we use $\Phi$ to define the deconfinement phase transition.

Making the mean field approximation and performing 
the path integral over quark field, 
one can obtain the thermodynamic potential $\Omega$ (per volume), 
\begin{align}
\Omega =& -2 N_f \int \frac{d^3{\rm p}}{(2\pi)^3}
         \Bigl[ 3 E ({\rm p}) \nonumber\\
        & + \frac{1}{\beta}
           \ln~ [1 + 3(\Phi+\Phi^{*} e^{-\beta E^-({\bf p})}) 
           e^{-\beta E^-({\bf p})}+ e^{-3\beta E^-({\bf p})}]
         \nonumber\\
        & + \frac{1}{\beta} 
           \ln~ [1 + 3(\Phi^{*}+{\Phi e^{-\beta E^+({\bf p})}}) 
              e^{-\beta E^+({\bf p})}+ e^{-3\beta E^+({\bf p})}]
	      \Bigl]\nonumber\\
        & +U_{\rm M}+{\cal U}. 
\label{eq:E12} 
\end{align}
where, $\sigma = \langle \bar{q}q \rangle $, 
$\Sigma_{\rm s} = -2 G_{\rm s} \sigma$, $M=m_0 + \Sigma_{\rm s}$, 
$U_{\rm M}= G_{\rm s} \sigma^2$, $E({\rm p})=\sqrt{{\bf p}^2+M^2}$ and 
$E^\pm({\rm p})=E({\rm p})\pm \mu =E({\rm p})\pm i\theta/\beta$. 
In \eqref{eq:E12}, only the first term of the right-hand side diverges. 
It is then regularized by the three-dimensional momentum
cutoff $\Lambda$~\cite{Fukushima,Ratti}.
We use ${\cal U}$ of Ref.~\cite{Rossner} that is fitted to a lattice QCD 
simulation 
in the pure gauge theory at finite $T$~\cite{Boyd,Kaczmarek}: 
\begin{align}
&{\cal U} = T^4 \Bigl[-\frac{a(T)}{2} {\Phi}^*\Phi\notag\\
      &~~~~~+ b(T)\ln(1 - 6{\Phi\Phi^*}  + 4(\Phi^3+{\Phi^*}^3)
            - 3(\Phi\Phi^*)^2 )\Bigr], \label{eq:E13}\\
&a(T)   = a_0 + a_1\Bigl(\frac{T_0}{T}\Bigr)
                 + a_2\Bigl(\frac{T_0}{T}\Bigr)^2,
 ~~~b(T)=b_3\Bigl(\frac{T_0}{T}\Bigr)^3  \label{eq:E14}
\end{align}
where parameters are summarized in Table I.  
The Polyakov potential yields a first-order deconfinement phase transition at 
$T=T_0$ in the pure gauge theory.
The original value of $T_0$ is $270$ MeV evaluated 
by the pure gauge lattice QCD calculation. 
However, the PNJL model with this value of $T_0$ yields somewhat larger value of the transition temperature at zero chemical potential than 
the full LQCD simulation~\cite{Karsch3,Karsch4,Kaczmarek2} predicts. 
Therefore, we rescale $T_0$ to 212~MeV; the detail will be shown in subsection 
\ref{No-mu}. 

\begin{table}[h]
\begin{center}
\begin{tabular}{llllll}
\hline
~~~~~$a_0$~~~~~&~~~~~$a_1$~~~~~&~~~~~$a_2$~~~~~&~~~~~$b_3$~~~~~
\\
\hline
~~~~3.51 &~~~~-2.47 &~~~~15.2 &~~~~-1.75\\
\hline
\end{tabular}
\caption{
Summary of the parameter set in the Polyakov sector
used in Ref.~\cite{Rossner}. 
All parameters are dimensionless. 
}
\end{center}
\end{table}

The variables $X=\Phi$, ${\Phi}^*$ and $\sigma$ 
satisfy the stationary conditions, 
\bea
\partial \Omega/\partial X=0. 
\label{eq:SC}
\label{condition}
\eea
The solutions of the stationary conditions do not give 
the global minimum $\Omega$ 
necessarily. There is a possibility 
that they yield a local minimum or even 
a maximum. We then have checked that the solutions yield 
the global minimum when the solutions $X(\theta)$ are inserted 
into (\ref{eq:E12}). 

The thermodynamic potential $\Omega$ of Eq. (\ref{eq:E12}) is not 
invariant under the ${\mathbb Z}_3$ transformation, 
\bea
\Phi(\theta) \to \Phi(\theta) e^{-i{2\pi k/{3}}} \;,\quad
\Phi(\theta)^{*} \to \Phi(\theta)^{*} e^{i{2\pi k/{3}}} \;, 
\eea
although 
${\cal U}$ of (\ref{eq:E13}) is invariant. 
Instead of the ${\mathbb Z}_3$ symmetry, however, 
$\Omega$ is invariant under the extended ${\mathbb Z}_3$ transformation, 
\begin{align}
&e^{\pm i \theta} \to e^{\pm i \theta} e^{\pm i{2\pi k\over{3}}},\quad  
\Phi(\theta)  \to \Phi(\theta) e^{-i{2\pi k\over{3}}}, 
\notag\\
&\Phi(\theta)^{*} \to \Phi(\theta)^{*} e^{i{2\pi k\over{3}}} .
\label{eq:K2}
\end{align}
This is easily understood as follows. 
It is convenient to introduce the modified Polyakov loop 
$\Psi \equiv e^{i\theta}\Phi$ and 
$\Psi^{*} \equiv e^{-i\theta}\Phi^{*}$ 
invariant under the transformation (\ref{eq:K2}). 
The extended ${\mathbb Z}_3$ transformation is then 
rewritten into 
\begin{align}
&e^{\pm i \theta} \to e^{\pm i \theta} e^{\pm i{2\pi k\over{3}}}, \quad
\Psi(\theta) \to \Psi(\theta), \notag\\ 
&\Psi(\theta)^{*} \to \Psi(\theta)^{*} ,
\label{eq:K2'}
\end{align}
and $\Omega$ is also into  
\begin{align}
\Omega = & -2 N_f \int \frac{d^3{\rm p}}{(2\pi)^3}
          \Bigl[ 3 E ({\rm p}) 
          + \frac{1}{\beta}\ln~ [1 + 3\Psi e^{-\beta E({\bf p})}
\notag\\
          &+ 3\Psi^{*}e^{-2\beta E({\bf p})}e^{\beta \mu_{\rm B}}
          + e^{-3\beta E({\bf p})}e^{\beta \mu_{\rm B}}]
\notag\\
          &+ \frac{1}{\beta} 
           \ln~ [1 + 3\Psi^{*} e^{-\beta E({\bf p})}
          + 3\Psi e^{-2\beta E({\bf p})}e^{-\beta\mu_{\rm B}}
\notag\\
          &+ e^{-3\beta E({\bf p})}e^{-\beta\mu_{\rm B}}]
	      \Bigl]+U_{\rm M}+ {\cal U} ,
\label{eq:K3} 
\end{align}
where $\beta\mu_{\rm B}=3 \beta\mu= 3 i \theta$. 
Obviously, $\Omega$ is invariant under the extended ${\mathbb Z}_3$ 
transformation \eqref{eq:K2'}, since 
it is a function of only extended ${\mathbb Z}_3$ invariant 
quantities, $e^{3i\theta}$ and $\tilde{X}(=\Psi, \Psi^{*},\sigma$). 
The explicit $\theta$ dependence appears only 
through the factor $e^{3i\theta}$ in (\ref{eq:K3}). Hence, 
the stationary conditions (\ref{eq:SC}) show that 
$\tilde{X}=\tilde{X}(e^{3i\theta})$. 
Inserting the solutions back to (\ref{eq:K3}), one can see that 
$\Omega=\Omega(e^{3i\theta})$. 
Thus, $\tilde{X}$ and $\Omega$ have the 
RW periodicity, 
\begin{align}
\tilde{X}(\theta +\frac{2\pi k}{3})=\tilde{X}(\theta), \quad {\rm and} \quad 
\Omega(\theta +\frac{2\pi k}{3})=\Omega(\theta), 
\end{align}
while the Polyakov loop $\Phi$ and its Hermitian conjugate $\Phi^*$ have the properties 
\begin{eqnarray}
\Phi (\theta +\frac{2\pi k}{3})&=&e^{-i2\pi k/3}\Phi (\theta ),
\nonumber\\
\Phi (\theta +\frac{2\pi k}{3})^*&=&e^{i2\pi k/3}\Phi (\theta )^*. 
\end{eqnarray}

\section{Numerical results}
\label{Numerical-results}

\subsection{Thermal system with no chemical potential}
\label{No-mu}

First, we consider the thermal system with no chemical potential to 
determine the parameters, $m_0$, $G_\mathrm{s}$, $\Lambda$ and $T_0$ 
of the PNJL model. 
In the lattice calculations~\cite{Karsch4,Karsch3,Kaczmarek2}, 
the pseudocritical temperature $T_\mathrm{c}(\sigma)$ 
of the crossover chiral phase transition 
coincides with that $T_\mathrm{c}(\Phi)$ of the crossover deconfinement one 
within $10$~\% error: 
$T_\mathrm{c}({\sigma}) \approx T_\mathrm{c} (\Phi) \approx 173\pm8$~MeV~\cite{Karsch4}. 

The parameter set, $\Lambda =631.5$~MeV, 
$G_{\rm s}=5.498$~[GeV$^{-2}]$ and $m_0=5.5$~MeV, 
can reproduce the pion decay constant $f_{\pi}=93.3$~MeV and the pion mass $M_{\pi}=138$~MeV at $T=\mu=0$~\cite{Kashiwa}, and keeps a good reproduction also at finite $T$~\cite{Rossner}. 
We then adopt these values for $\Lambda$, 
$G_{\rm s}$ and $m_0$. 
We adjust $T_0$ so that the PNJL calculation can reproduce the lattice result $T_\mathrm{c} (\Phi)=173$~MeV; the value is $T_0=$212~MeV. 
The parameter set thus determined is shown as set A in Table~\ref{PNJL-parameter}. 

\begin{table}[h]
\begin{center}
\begin{tabular}{cccc}
\hline
~~set~~&~~$G_{\rm s}$~~&~~$G_{\rm s8}$~~&~~$G_{\rm v}$~~\\
\hline
~~A~~&~~5.498${\rm GeV}^{-2}$~~&~~0~~&~~0~~\\
\hline
~~B~~&~~4.673${\rm GeV}^{-2}$~~&~~452.12${\rm GeV}^{-8}$~~&~~0~~\\
\hline
~~C~~&~~4.673${\rm GeV}^{-2}$~~&~~452.12${\rm GeV}^{-8}$~~&~~4.673${\rm GeV}^{-2}$~~\\
\hline
\end{tabular}
\caption{
Summary of the parameter sets in the PNJL calculations.
The parameters $\Lambda$, $m_0$ and $T_0$ are common among the three sets; 
$\Lambda =631.5$~MeV, $m_0=5.5$~MeV and $T_0=212$~MeV. 
}
\label{PNJL-parameter}
\end{center}
\end{table}

\begin{figure}[htbp]
\begin{center}
 \includegraphics[width=0.30\textwidth,angle=-90]{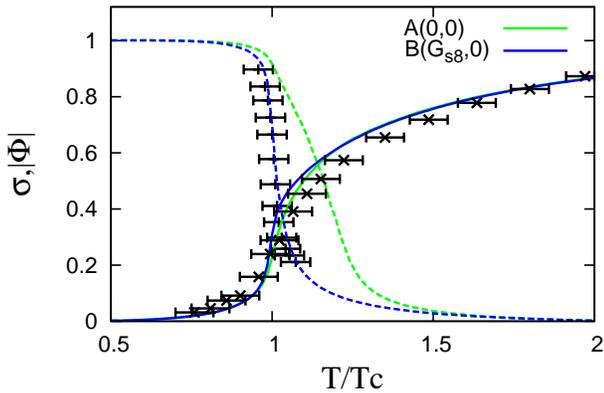}
\end{center}
\caption{Chiral condensate $\sigma$ normalized by
 $\sigma(T=0, \mu=0)$ and the absolute value of the Polyakov loop $\Phi$. 
 The blue (green) curve represents the PNJL result 
 of parameter set B (A) with (without) 
 the scalar-type eight-quark interaction; 
 $\sigma$ ($|\Phi|$) is denoted by the dashed (solid) curve. 
 Lattice data ($+$) 
 on $\sigma$ are taken from Ref.~\cite{Karsch3} and those ($\times$) 
 on $|\Phi|$ are  from Ref.~\cite{Kaczmarek2}. 
  The lattice data 
 are plotted with 10 \% error bar, since lattice calculations have 
 10 \% error in determining $T_\mathrm{c}$~\cite{Karsch4}. 
  }
\label{T-dep-Nf2-mu0}
\end{figure}

Figure \ref{T-dep-Nf2-mu0} shows the chiral condensate $\sigma$ normalized 
by $\sigma_0=\sigma|_{T=0, \mu=0}$ and the absolute value of 
the Polyakov loop $\Phi$ as a function of $T/T_\mathrm{c}$. 
In this paper $T_\mathrm{c}$ is always taken to be 173~MeV. 
The green curves represent the PNJL results of parameter set A, where 
$\sigma_0=-0.0302$~[GeV$^3]$ in this case. 
Lattice QCD data~\cite{Karsch4,Karsch3,Kaczmarek2} are also plotted 
by cross symbols with 10~\% error bar; 
$\sigma$ and $|\Phi|$ measured  as a function of $T/T_\mathrm{c}$ 
in Refs.~\cite{Karsch4,Karsch3,Kaczmarek2} have only small errors, but 
we have added 10 \% error 
that the lattice calculation~\cite{Karsch4} has in determining $T_\mathrm{c}$. 
For $|\Phi|$ the PNJL result 
(green solid curve) reasonably agrees with the lattice one ($\times$). 
For $\sigma$, however, the PNJL result (green dashed curve) considerably 
overshoots the lattice data ($+$). 

\begin{figure}[htbp]
\begin{center}
 \includegraphics[width=0.27\textwidth,angle=-90]{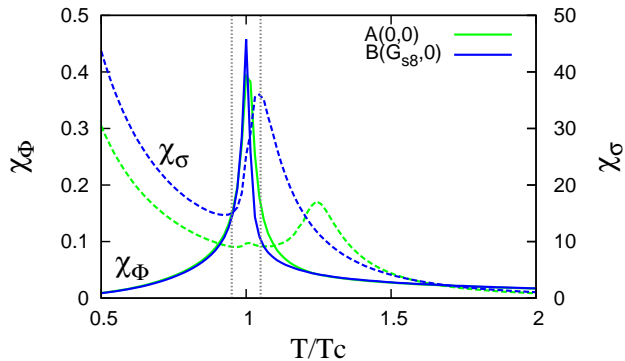}
\end{center}
\caption{$T$ dependence of 
chiral and Polyakov-loop susceptibilities, $\chi_{\sigma}$ (right scale) and 
 $\chi_{\Phi}$ (left scale). 
 The blue (green) curve represents the PNJL result 
 of parameter set B (A) 
 with (without) the scalar-type eight-quark interaction;
 $\chi_{\sigma}$ ($\chi_{\Phi}$) is denoted by the dashed (solid) curve. 
 The region between two vertical gray lines $T=(1 \pm 0.05)T_\mathrm{c}$ is the 
 prediction of lattice calculations~\cite{Karsch4}. 
   }
\label{suscept-Nf2-mu0}
\end{figure}

Figure \ref{suscept-Nf2-mu0} 
represents results of the PNJL calculations for 
chiral and Polyakov-loop susceptibilities, 
$\chi_{\sigma}$ and $\chi_{\Phi}$~\cite{Fukushima2}. 
Peak positions of $\chi_{\sigma}$ and $\chi_{\Phi}$ 
show $T_\mathrm{c}({\sigma})$ and $T_\mathrm{c} (\Phi)$, respectively. 
The PNJL results (green curves) 
of parameter set A give $T_\mathrm{c}({\sigma})/T_\mathrm{c}=1.25$ and 
$T_\mathrm{c} (\Phi)/T_\mathrm{c}=1$, while the lattice simulations yield 
$T_\mathrm{c}({\sigma})/T_\mathrm{c}=1 \pm 0.05$ and 
$T_\mathrm{c} (\Phi)/T_\mathrm{c}=1 \pm 0.05$. 
The PNJL results are consistent with 
the lattice ones for $T_\mathrm{c}(\Phi)$, but not for $T_\mathrm{c}({\sigma})$.

Now we introduce the scalar-type eight-quark interaction~\cite{Kashiwa}, 
\begin{align}
G_{\rm s8}[(\bar{q}q)^2+(\bar{q}i\gamma_5\vec{\tau}q)^2]^2 ,
\label{eq:E1a}
\end{align}
since the difference $T_\mathrm{c}({\sigma})-T_\mathrm{c}(\Phi)$ is reduced 
by the interaction~\cite{Sakai}. 

Since $f_{\pi}$ and $M_{\pi}$ calculated with PNJL 
depend on the strength of $G_{\rm s8}$, for 
each value of $G_{\rm s8}$ the strength of $G_{\rm s}$ is re-adjusted 
so as to reproduce the measured values $f_{\pi}=93.3$~MeV 
and $M_{\pi}=138$~MeV. As $G_{\rm s8}$ increases from zero, 
$T_\mathrm{c}({\sigma})$ calculated with PNJL decreases toward $T_\mathrm{c}=173$~MeV. 
When $G_{\rm s8}=452.12$~GeV$^{-8}$, 
the ratio $T_\mathrm{c}({\sigma})/T_\mathrm{c}$ becomes 1.05 and hence 
consistent with 
the corresponding lattice result within $10~\%$ error. 
We adopt this strength. This parameter set is shown 
as set B in Table~\ref{PNJL-parameter}. 
As shown in Fig.~\ref{T-dep-Nf2-mu0}, 
the PNJL results (blue curves) of parameter set B well reproduce 
the lattice results for both the chiral condensate and the Polyakov loop.  

\subsection{Thermal system with imaginary chemical potential}
\label{finite-imaginary-mu-B}

In this subsection, we consider the thermal system with 
finite imaginary chemical potential and 
compare the PNJL result with the lattice data~\cite{FP}(\cite{Chen}) 
in which the lattice size is $8^3 \times 4$ and 
the two-flavor KS(Wilson) fermion is considered.  

\begin{figure}[htbp]
\begin{center}
 \includegraphics[width=0.27\textwidth,angle=-90]{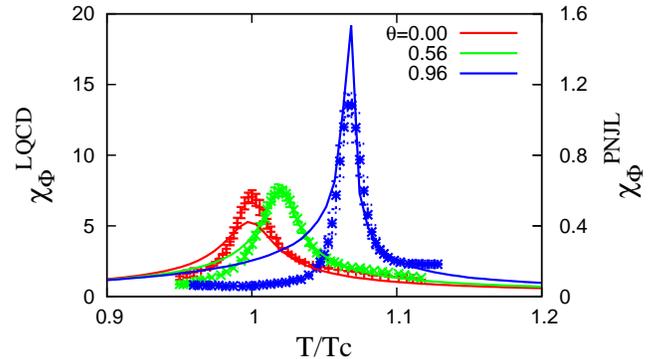}
 \end{center}
\caption{$T$ dependence of the Polyakov-loop susceptibilities 
in three cases of $\theta=0,~0.56$ and 0.96; each case is distinguished by 
using different colors. 
The solid curves represent the PNJL results of set A (right scale). 
Lattice data shown by crosses (left scale) are taken from Ref.~\cite{Chen}.}
\label{fig-Pol-suscep}
\end{figure}

First, we analyze the deconfinement phase transition. 
Since the eight-quark interaction hardly changes the 
Polyakov loop, 
we do the PNJL calculation with parameter set A. 
Figure~\ref{fig-Pol-suscep} presents $T$ dependence of 
the Polyakov-loop susceptibility $\chi_{\Phi}$ in three cases of 
$\theta=0,~0.56$ and 0.96; each case is distinguished by 
using different colors. For each $\theta$, the PNJL result (solid curve) 
reproduces the corresponding lattice result (crosses) in its peak position. 
Thus, the PNJL results are consistent with the lattice ones 
for the pseudocritical temperature of 
the crossover deconfinement phase transition. 

\begin{figure}[htbp]
\begin{center}
 \includegraphics[width=0.30\textwidth,angle=-90]{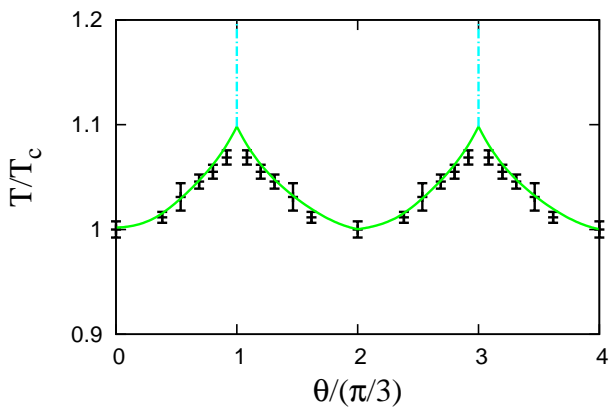}
 \end{center}
\caption{Phase diagram on the $\theta$--$T$ plane. The solid curve represents 
the deconfinement phase transition, while the dot-dashed line does 
the RW phase transition predicted by the PNJL calculation with set A. 
Lattice data are taken from Ref.~\cite{Chen}.}
\label{fig-PD-pol}
\end{figure}

\begin{figure}[htbp]
\begin{center}
 \includegraphics[width=0.30\textwidth,angle=-90]{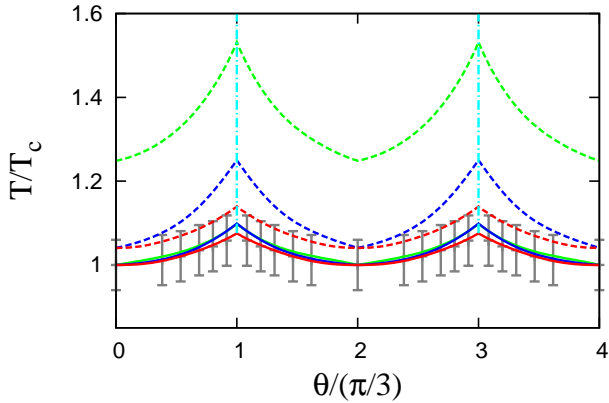}
 \end{center}
\caption{Phase diagrams of the chiral phase transition in the 
imaginary chemical potential region calculated with three parameter sets are presented by 
dashed curves;  
green, blue and red ones are results of the PNJL calculations 
with set A, B and C, respectively. 
Lattice data~\cite{Chen} are shown with 10\% error that 
$T_\mathrm{c}$ has~\cite{Karsch4}.
The deconfinement phase transition curves (solid curves) are also shown for comparison. 
} 

\label{PD-imaginary-mu}
\end{figure}

\begin{figure}[htbp]
\begin{center}
 \includegraphics[width=0.30\textwidth,angle=-90]{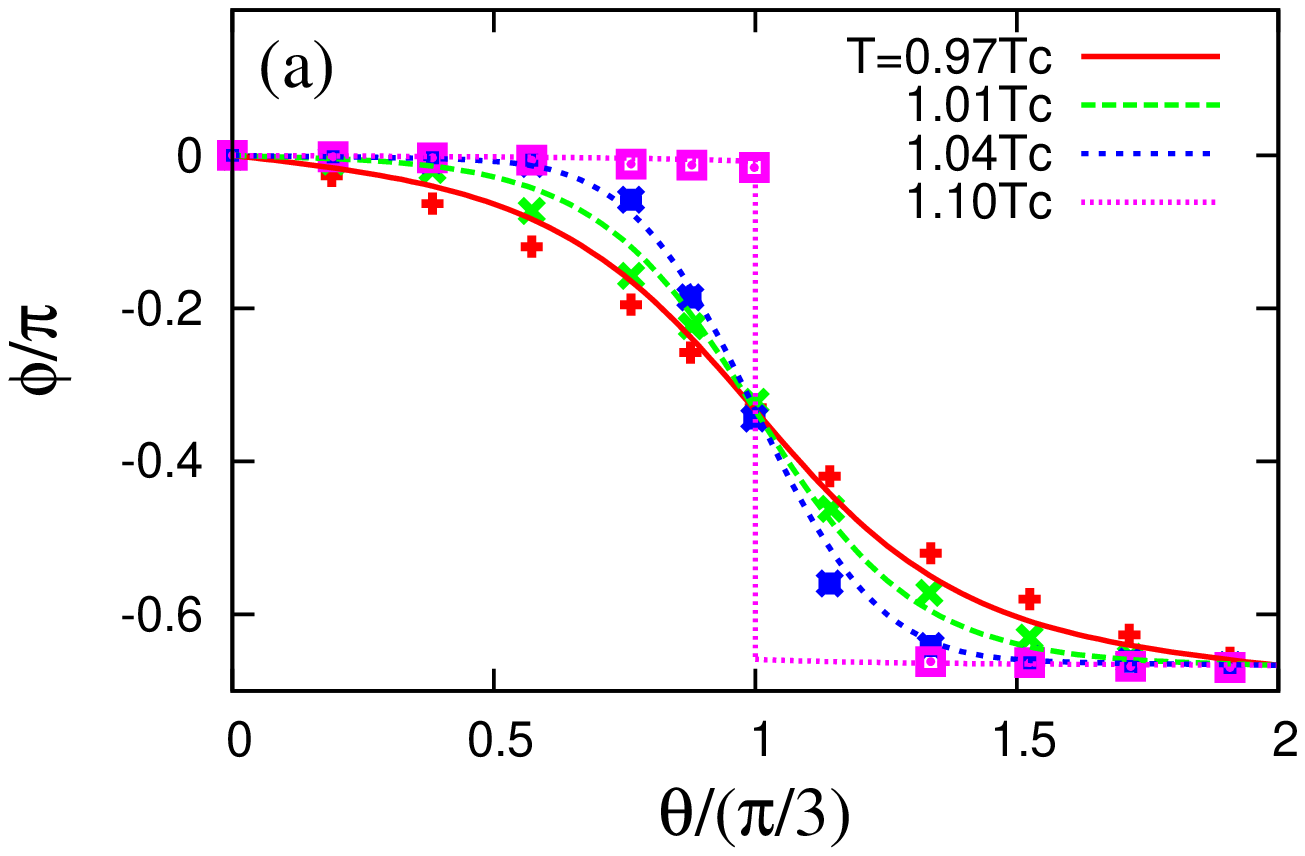}
 \includegraphics[width=0.30\textwidth,angle=-90]{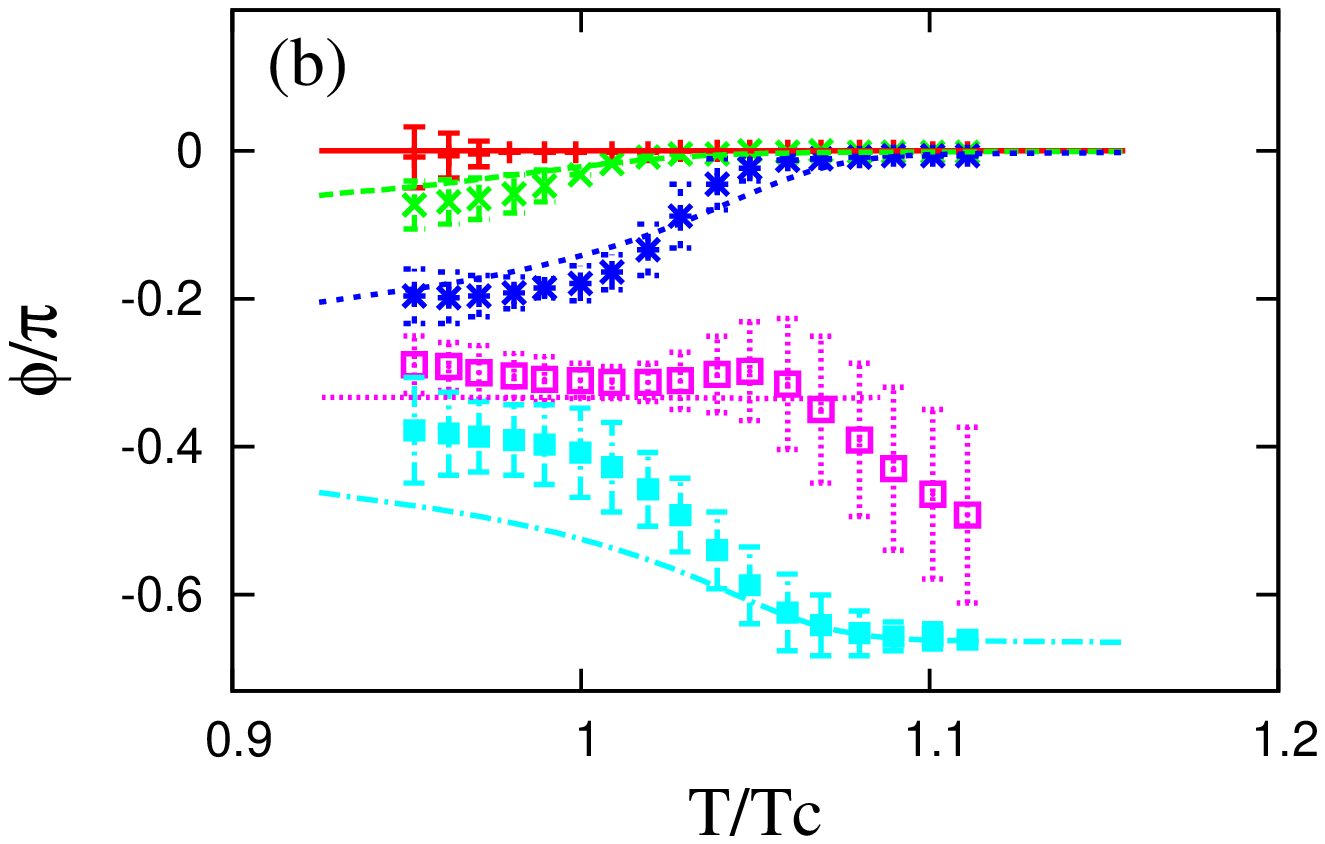}
\end{center}
\caption{Phase $\phi$ of the Polyakov loop as a function of 
(a) $\theta$ and (b) $T$. Lattice data~\cite{Chen,FP} 
are plotted by symbols. 
Curves represent results of PNJL calculations with set A. 
In panel (b), five cases (red, green, blue, pink and light blue) 
from top to bottom represent results of 
$\theta/(\pi/3)=0,~0.4,~0.8,~1.0$ and 1.2, respectively. 
The pink line terminates at 
$T=T_{\rm RW}=1.09~T_\mathrm{c}$, since $\phi$ is singular at $T > T_{\rm RW}$ 
in the case of $\theta=\pi/3$. 
}
\label{fig-Pol-phase}
\end{figure}

Figure~\ref{fig-PD-pol} presents the phase diagram of the deconfinement phase 
transition in the $\theta$-$T$ plane, where $\theta$ 
is divided by $\pi/3$ and $T$ is normalized by $T_\mathrm{c}=173$~MeV. 
Lattice data~\cite{Chen} measured as a function of $T/T_\mathrm{c}$ have 
only small errors, as shown by thick error bars in Fig.~\ref{fig-PD-pol}. 
This is an error bar in the case that lattice calculations have no error 
in $T_\mathrm{c}$. However, 
the lattice calculation~\cite{Karsch4} has about 10~\% error 
in determining $T_\mathrm{c}$, as mentioned in subsection \ref{No-mu}.  
This 10~\% error should be added to the original small error; 
this 10~\% error will be shown later in Fig. 
\ref{PD-imaginary-mu}. 
The PNJL result (solid curve) of set A agrees with the lattice one 
(crosses) within the error bars. 
The phase diagram has a periodicity of $2\pi/3$ in $\theta$. 
This is called the Roberge and Weiss (RW) periodicity~\cite{RW}. 
The phase diagram is also $\theta$ even, because so is $\chi_{\Phi}$. 
On the dot-dashed line going up from an endpoint 
$(\theta_{\rm RW},T_{\rm RW})=(\pi/3,1.09T_\mathrm{c})$, 
the quark number density $n$ 
and the phase $\phi$ of the Polyakov loop are discontinuous in 
the PNJL calculations~\cite{Sakai,Sakai1}. 
This is called the RW phase transition line. 
The lattice data~\cite{Chen,FP} on $\phi$ are also discontinuous on the line, 
as shown later in Fig.\ref{fig-Pol-phase}. 
Thus, the PNJL result is consistent with the lattice results~\cite{Chen,FP} 
also for the location of the RW phase transition line. 

\begin{figure}[htbp]
\begin{center} 
 \includegraphics[width=0.27\textwidth,angle=-90]{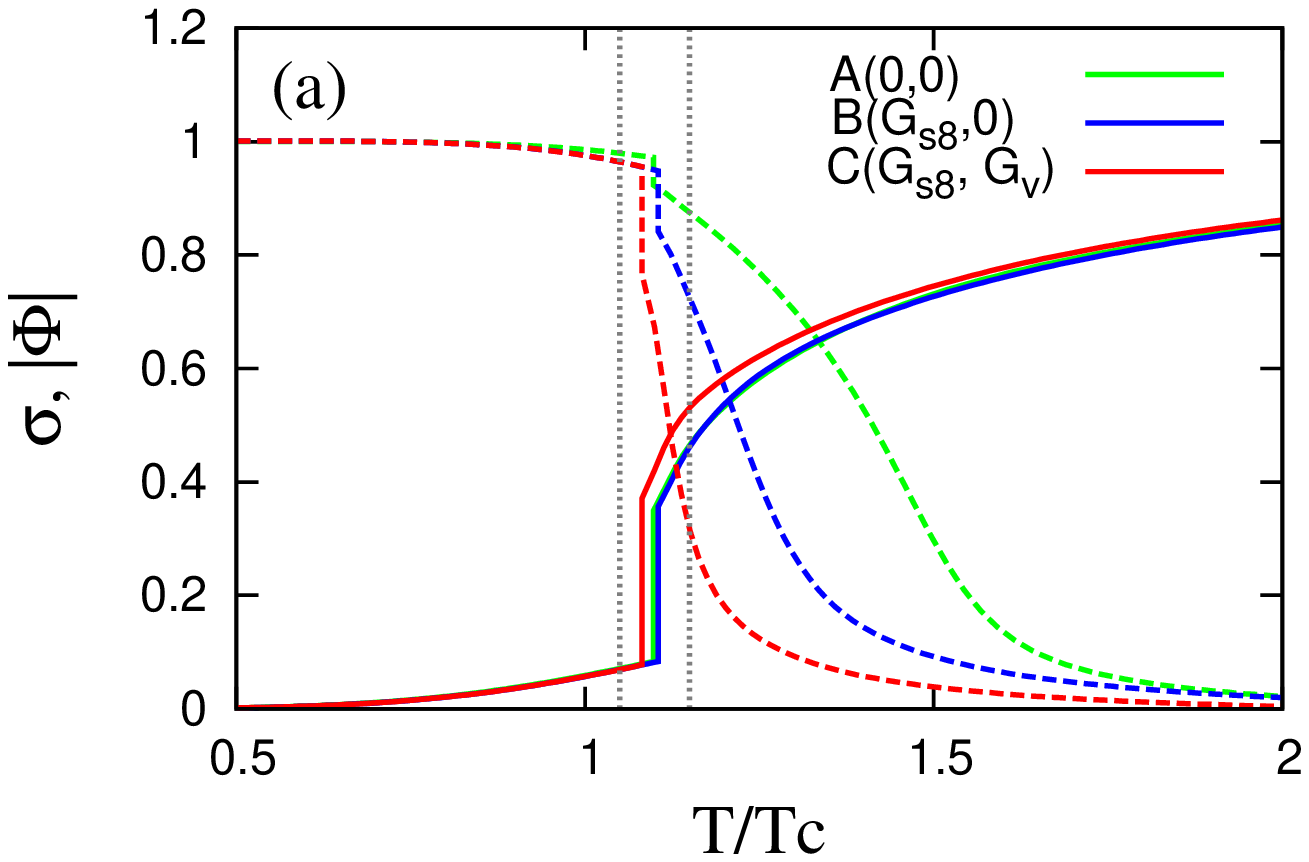}
 \includegraphics[width=0.27\textwidth,angle=-90]{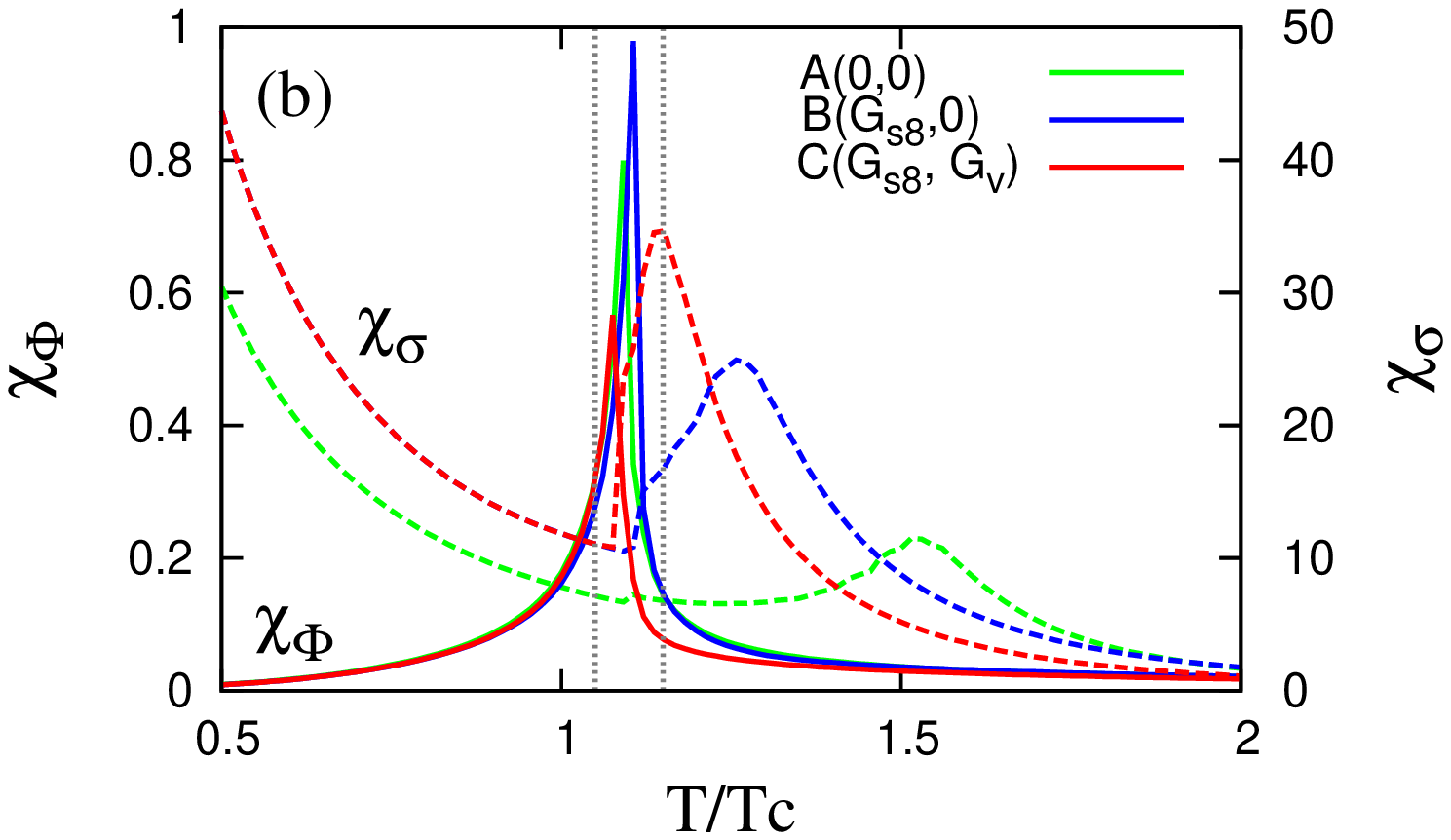}
\end{center}
\caption{$T$ dependence of (a) the normalized chiral condensate 
and the absolute value of the Polyakov loop and 
(b) the susceptibilities $\chi_{\sigma}$ (right scale) and $\chi_{\Phi}$ (left scale) 
at $\theta=\pi/3$. 
In panel (a), $\sigma$ ($|\Phi|$) is denoted by the dashed (solid) curve. 
In panel (b), $\chi_{\sigma}$ ($\chi_{\Phi}$) is 
denoted by the dashed (solid) curve. 
The PNJL calculations are done with three parameter sets of A, B and C and 
these are distinguished by using different colors, 
green, blue and red, respectively.}
\label{suscept-imaginary-mu}
\end{figure}

The lattice simulations~\cite{Chen,FP} point out that 
$T_\mathrm{c}(\sigma)$ agrees with  $T_\mathrm{c}(\Phi)$ within numerical errors 
in the entire region $0 \le \theta \le 2\pi/3$. 
We then take the case of $\theta=\pi/3$ to consider this point. 
It is predicted by the lattice simulations 
that $T_\mathrm{c}(\sigma)$ and $T_\mathrm{c}(\Phi)$ are located in 
the region between two vertical gray lines of Fig.~\ref{suscept-imaginary-mu}.
Panel (a) shows $\sigma$ and $|\Phi|$ as a function $T/T_\mathrm{c}$ and panel (b)
does $\chi_{\sigma}$ and $\chi_{\Phi}$ as a function $T/T_\mathrm{c}$. 
The green (blue) curves represent results of the PNJL calculations 
with set A (B). 
The eight-quark interaction hardly shifts the peak position of 
$\chi_{\Phi}$, i.e. $T_\mathrm{c}(\Phi)$, from the value $1.09T_\mathrm{c}$. 
The peak position is consistent with 
the lattice result shown by the region between two vertical gray lines. 
In contrast, the eight-quark interaction largely shifts 
the peak position of $\chi_{\sigma}$, i.e. $T_\mathrm{c}(\sigma)$,  
from $1.53T_\mathrm{c}$ to $1.24T_\mathrm{c}$, but the shifted value still deviates from 
$T_\mathrm{c}(\Phi)=(1.1\pm0.05)T_\mathrm{c}$, that is, the lattice data near $\theta=\pi/3$~\cite{Chen,FP} 
shown by the region between two vertical gray lines. 

In order to solve this problem, 
we introduce the vector-type four-quark interaction 
\bea
-G_{\rm v}({\bar q}\gamma_\mu q)^2
\eea
and add it to the PNJL Lagrangian ${\cal L}$; see 
Ref.~\cite{Sakai1} for the detail of this formulation. 
As mentioned in Ref.~\cite{Sakai1}, 
the phase structure in the real chemical potential region is quite 
sensitive to the strength of the coupling $G_{\rm v}$. 
It is then important to determine the strength, but 
it has not been done yet. Since the vector-type interaction does not change 
the pion mass and the pion decay constant at $T=\mu=0$ and the chiral 
condensate and the Polyakov loop at $T \ge 0$ and $\mu=0$, 
we can simply add the interaction to set B. 
As $G_{\rm v}$ increases from zero, $T_\mathrm{c}(\sigma)$ goes down toward 
$T_\mathrm{c}(\Phi)$, while $T_\mathrm{c}(\Phi)$ moves little. 
When $G_{\rm v}=4.673~$GeV$^{-2}$, 
$T_\mathrm{c}(\sigma)$ gets into the region between the vertical gray lines. 
We adopt this strength of $G_{\rm v}$. This set is shown as set C 
in Table \ref{PNJL-parameter}. 

Figure~\ref{PD-imaginary-mu} shows the phase diagram 
of the chiral phase transition determined by $T_\mathrm{c}(\sigma)$. 
Green, blue and red curves are results of the PNJL calculations 
with sets A, B and C, respectively. 
In the entire region $0 \le \theta \le 2\pi/3$, the eight-quark interaction 
moves $T_\mathrm{c}(\sigma)$ down 
from the green dashed curve (set A) to the blue one (set B). 
However, the blue dashed curve still overshoots the lattice result (symbols) with 10 \% error near $\theta=\pi/3$. 
The vector-type interaction makes the blue dashed curve go down to the red one (set C) that is consistent with the lattice result~\cite{Chen}. 
Thus, the PNJL calculations with set C can reproduce 
the lattice result~\cite{Chen,FP} 
that $T_\mathrm{c}(\sigma)$ coincides with $T_\mathrm{c}(\Phi)$ within numerical errors in the entire region $0 \le \theta \le 2\pi/3$.

Figure \ref{fig-Pol-phase}(a) shows 
$\theta$ dependence of the phase $\phi$ of 
$\Phi$ for four cases of $T/T_\mathrm{c}=0.97,~1.01,~1.04$ and 1.10; 
each case is distinguished by using different colors. 
The PNJL results (curves) well simulate the lattice data~\cite{Chen,FP}
 (symbols). 
It is found from both the results that 
$\phi$ is continuous at $\theta=\pi/3$ 
in the low-$T$ side $T \leq T_{\mathrm{RW}}=1.09T_\mathrm{c}$, 
but it is discontinuous at $\theta=\pi/3$ in the 
high-$T$ side $T > T_{\mathrm{RW}}$. 
Hence, the RW phase transition takes place at $T > T_{\mathrm{RW}}=1.09T_\mathrm{c}$ 
and $\theta=\pi/3$. 

Figure \ref{fig-Pol-phase}(b) shows 
$T$ dependence of $\phi$ for five cases of $\theta/(\pi/3)=0,~0.4,~0.8,~1.0$ 
and 1.2. The PNJL results (curves) well reproduce 
the lattice data~\cite{Chen,FP} (symbols). 
For $\theta < \pi/3$ the phase $\phi$ tends to zero 
as $T$ increases, while for $\theta > \pi/3$ 
it does to $-2\pi/3$ as $T$ increases. 
When $\theta=\pi/3$, the RW phase transition takes place 
at $T > T_{\mathrm{RW}}=1.09T_\mathrm{c}$ and then the phase $\phi$ is singular there, 
so that the pink line terminates at $T=T_{\mathrm{RW}}$. 
In the high-$T$ limit, the region (I) $-\pi/3<\theta<\pi/3$ 
has $\phi=0$ and the region (II) $\pi/3<\theta<\pi$ does $\phi=-2\pi/3$. 
Thus, the region (II) is a ${\mathbb Z}_3$ image 
of the region (I), and the region (III) $\pi<\theta<5\pi/3$ is another 
${\mathbb Z}_3$ image of the region (I). 

\subsection{Thermal system with real chemical potential}
\label{finite-imaginary-mu-C}

\begin{figure}[htbp]
\begin{center} 
 \includegraphics[width=0.30\textwidth,angle=-90]{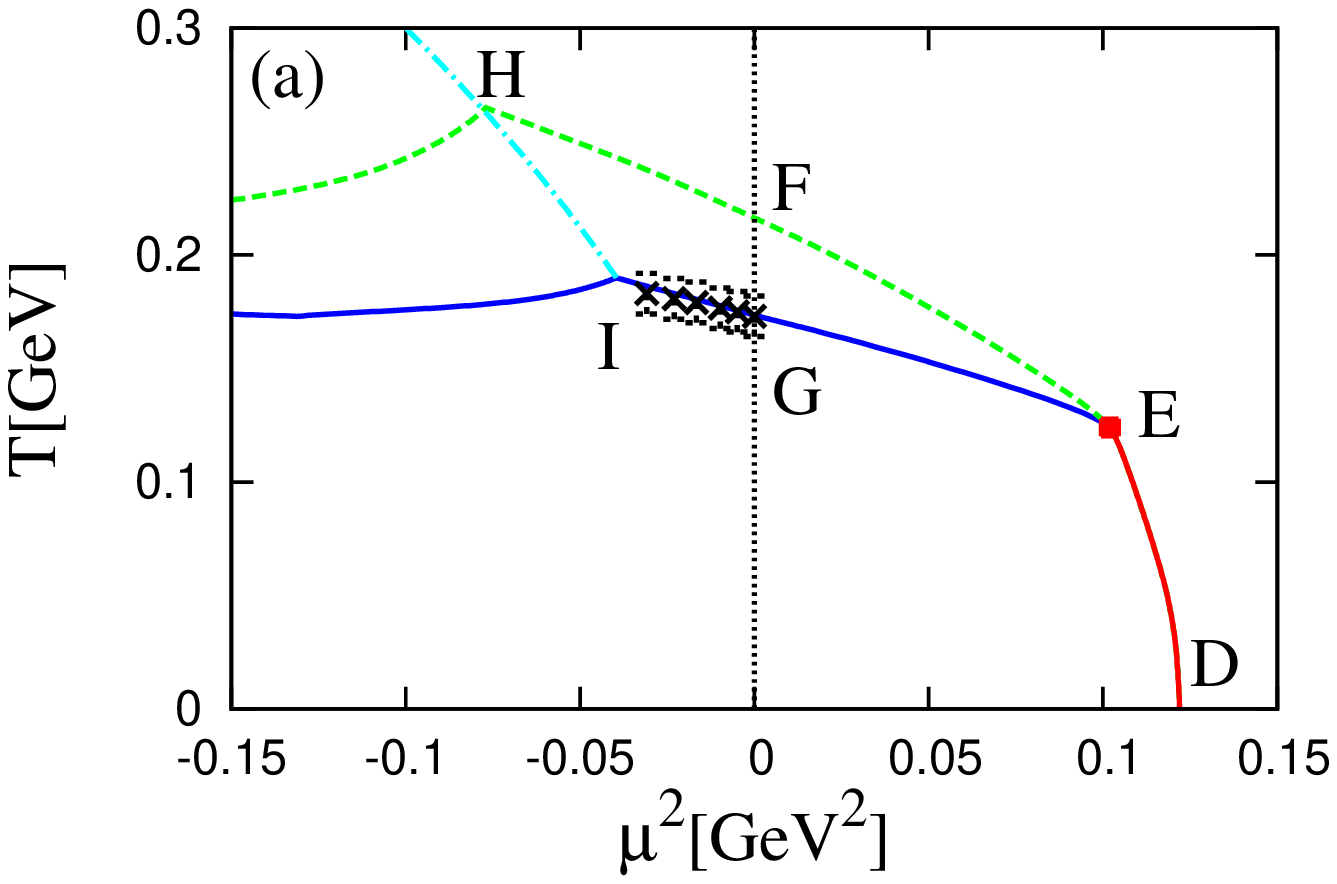}
 \includegraphics[width=0.30\textwidth,angle=-90]{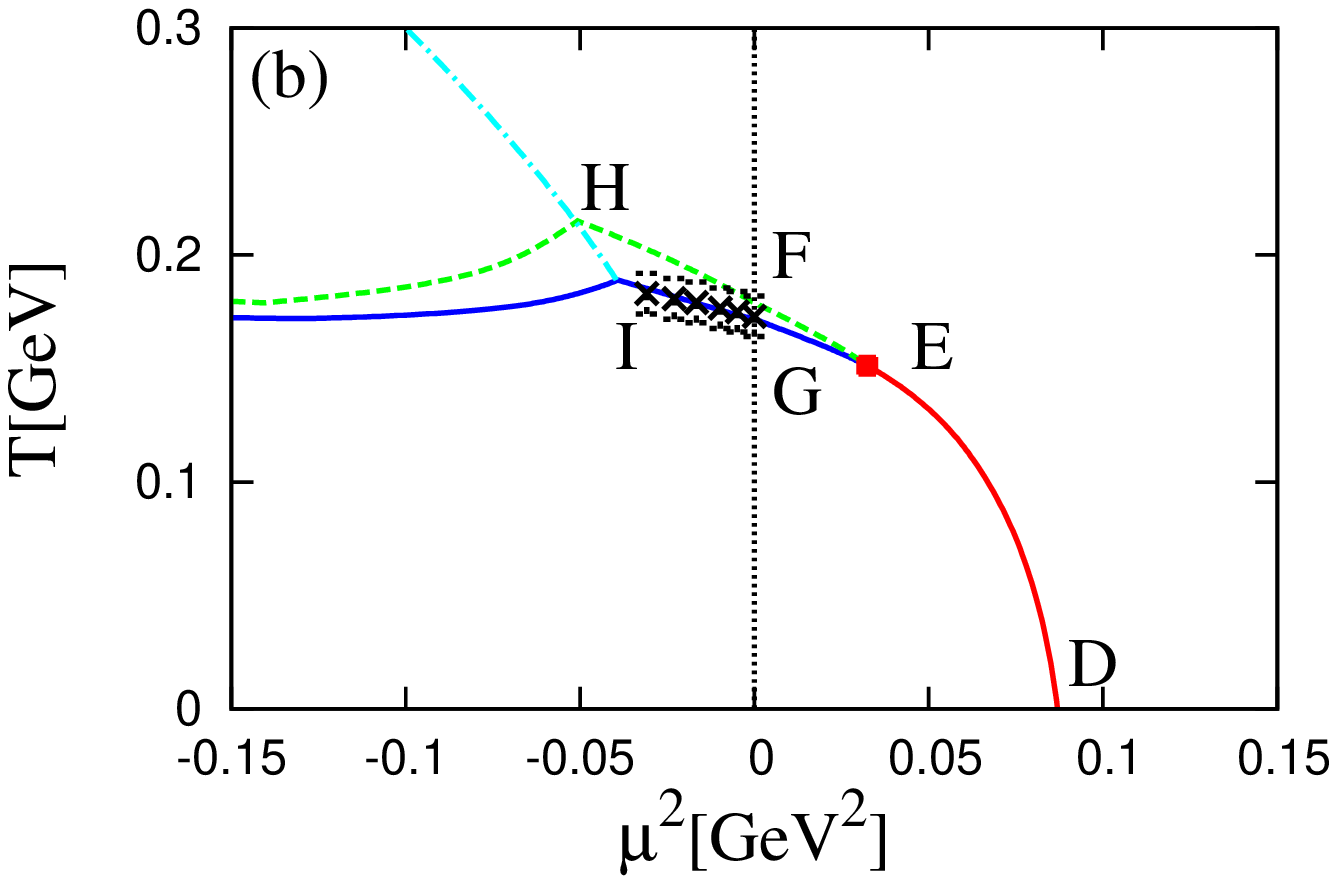}
 \includegraphics[width=0.30\textwidth,angle=-90]{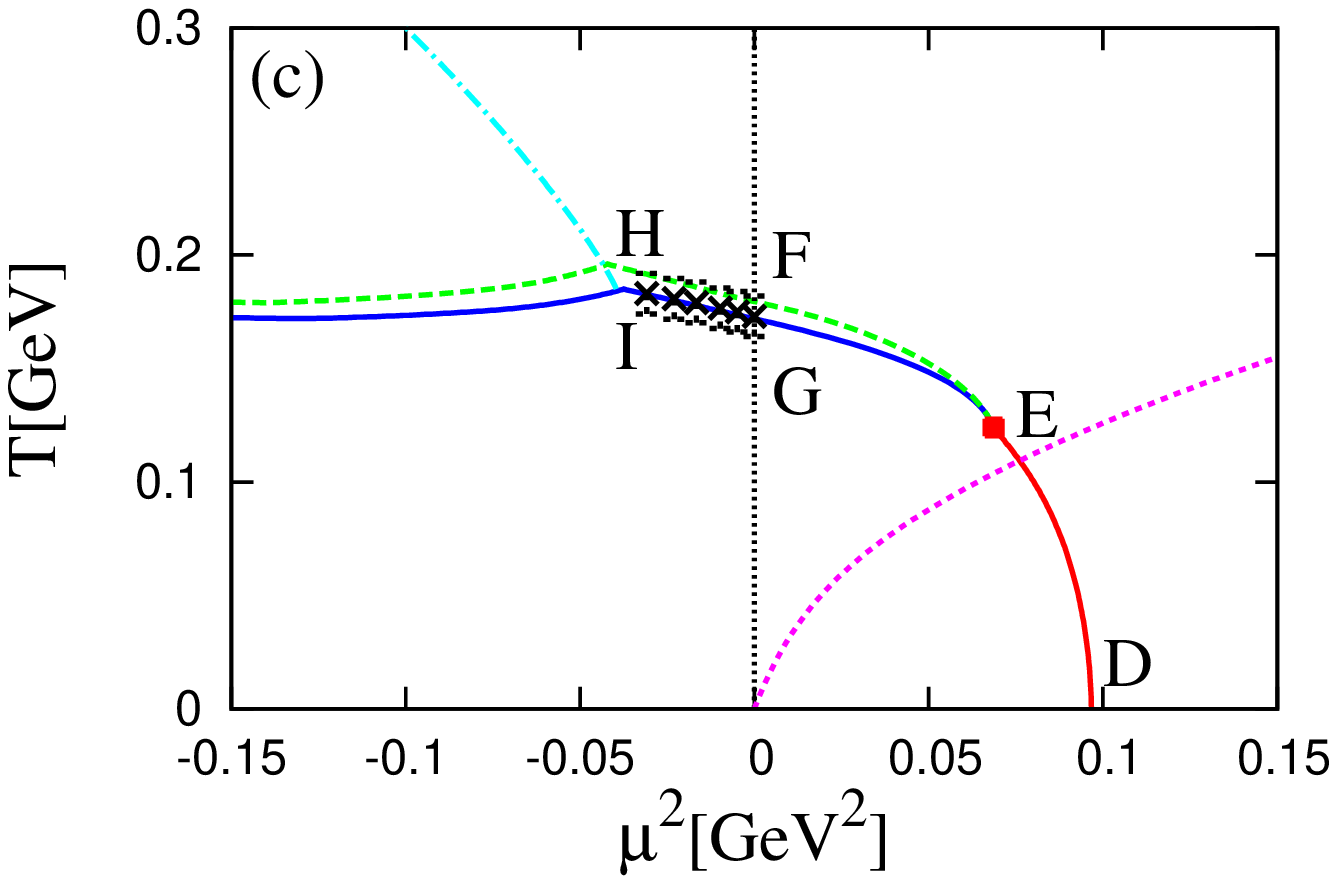}
\end{center}
\caption{Phase diagram in the real chemical potential region. 
(a), (b), and (c) are calculated with the parameter set A, B, and C, 
respectively. Cross symbols with error bars indicate the latticle data 
taken from Ref.~\cite{Chen}. Points D--I are explained in the text. }
\label{phase-diagram-Nf2-real-mu}
\end{figure}

In this subsection, we predict the phase diagram in the 
real $\mu$ region by using the PNJL model. 
In Fig.~\ref{phase-diagram-Nf2-real-mu}, panels (a)-(c) represent 
results of the PNJL calculations with sets A, B and C, respectively. 
Panel (c) is the most reliable result, since the PNJL result of set C is 
consistent with the lattice result~\cite{FP,Chen} in 
the imaginary chemical potential region. 
Comparing the three panels, we 
find that the vector-type four-quark interaction and the scalar-type 
eight-quark interaction give sizable effects on the phase structure. 
In particular for the critical endpoint E, the eight-quark interaction 
shifts point E to larger $T$ and smaller $\mu$, and the 
vector-type interaction moves it in the opposite direction. 
On the red solid curve between point E and point D 
both the first-order chiral and deconfinement phase transitions take place 
simultaneously. The light-blue dot-dashed curve moving up from point I represents 
the RW phase transition of first order, and point I is the critical endpoint. 
The green dashed curve between point H and point E means the crossover chiral 
phase transition and the blue solid curve between point I and point E does 
the crossover deconfinement phase transition. Point F (G) is 
a crossing point between the dashed (solid) curve and the $\mu=0$ line. 
Positions of points D--I are summarized in Table~\ref{points}. 
In panel (c), the pink dotted curve  represents 
the lower bound of the location $\mu_{\rm E}/T_{\rm E}$ of 
the critical endpoint E that the LQCD analyses of 
Ref.~\cite{Ejiri} predict. 
The position of point E in the case of parameter set C is consistent 
with the results of the LQCD analyses. 

\begin{table*}
\begin{center}
\begin{tabular}{ccccccc}
\hline
~~set~~&~~D~~&~~E~~&~~F~~&~~G~~&~~H~~&~~I~~\\
\hline
~~A~~&~~(2.02, 0.00)~~&~~(1.84, 0.72)~~&~~(0.00, 1.25)~~&~~(0.00, 1.00)
~~&~~($i\pi/3\times 1.53$, 1.53)~~&~~($i\pi/3\times 1.09$, 1.09)~~\\
\hline
~~B~~&~~(1.68, 0.00)~~&~~(1.02, 0.87)~~&~~(0.00, 1.05)~~&~~(0.00, 1.00)
~~&~~($i\pi/3\times 1.24$, 1.24)~~&~~($i\pi/3\times 1.09$, 1.09)~~\\
\hline
~~C~~&~~(1.80, 0.00)~~&~~(1.51, 0.72)~~&~~(0.00, 1.05)~~&~~(0.00, 1.00)
~~&~~($i\pi/3\times 1.13$, 1.13)~~&~~($i\pi/3\times 1.07$, 1.07)~~\\
\hline
\end{tabular}
\caption{Positions of points D-I in $\mu$-$T$ plane. 
The positions of these points are normalized as 
$(\mu/T_{\rm c},T/T_{\rm c})$ with $T_{\rm c}=173$~MeV.},
\label{points}
\end{center}
\end{table*}

\section{Summary}
\label{Summary}

We have tested the reliability of the PNJL model, 
comparing the model result with 
lattice data in the imaginary chemical potential ($\mu_\mathrm{I}=T \theta$) 
region. In this test, 
the model parameters except $G_{\rm v}$ are adjusted so as to reproduce 
the measured pion mass and decay constant at $T=\mu=0$ 
and lattice data~\cite{Karsch4,Karsch3,Kaczmarek2} at $T > 0$ and $\mu=0$. 
In this step the eight-quark interaction plays an important role to 
make $T_\mathrm{c}(\sigma)$ closer to $T_\mathrm{c}(\Phi)$ as discussed in 
our previous work~\cite{Sakai}. With the aid of this, 
the PNJL calculation with the eight-quark interaction but without 
the vector-type interaction well reproduces 
the lattice data~\cite{FP,Chen} at finite $\theta$ 
on $\Phi$ and $T_\mathrm{c}(\Phi)$, but not 
on $T_\mathrm{c}(\sigma)$ particularly near $\theta=\pi/3$ fully. 
The strength of $G_{\rm v}$ is then fitted so as to reproduce the data 
on $T_\mathrm{c}(\sigma)$ near $\theta=\pi/3$. 
The primary result of the lattice simulations is that 
$T_\mathrm{c}(\Phi)$ coincides with $T_\mathrm{c}(\sigma)$, 
within numerical errors, 
in the entire region of $\theta$~\cite{FP,Chen}. 
The PNJL model with the eight-quark and vector-type interactions 
can reproduce this property. 
Therefore, we can expect that the PNJL model with this parameter set 
is reliable also in the $\mu_\mathrm{R}$ region.

Finally, we quantitatively predict the phase diagram in the $\mu_\mathrm{R}$ 
region by using the PNJL model with the parameter set mentioned above. 
The critical endpoint does not disappear in virtue of 
the eight-quark interaction, even if the vector-type 
interaction is taken into account. 
This is the primary result of the present work. 
The lattice calculations at nonzero $\mu_\mathrm{I}$ have 
small lattice size ($8^3\times4$)~\cite{FP,Chen}. 
Therefore, it is highly expected that 
lattice simulations with larger size will be done in the $\mu_\mathrm{I}$ region.

\noindent
\begin{acknowledgments}
The authors thank A. Nakamura for useful discussions and suggestions. 
K.K. and H.K. also thank M. Imachi, H. Yoneyama and M. Tachibana for useful discussions. 
This work has been supported in part 
by the Grants-in-Aid for Scientific Research 
(18540280) of Education, Science, Sports, and Culture of Japan.
\end{acknowledgments}



\begin{thebibliography}{19}
\expandafter\ifx\csname natexlab\endcsname\relax\def\natexlab#1{#1}\fi
\expandafter\ifx\csname bibnamefont\endcsname\relax
  \def\bibnamefont#1{#1}\fi
\expandafter\ifx\csname bibfnamefont\endcsname\relax
  \def\bibfnamefont#1{#1}\fi
\expandafter\ifx\csname citenamefont\endcsname\relax
  \def\citenamefont#1{#1}\fi
\expandafter\ifx\csname url\endcsname\relax
  \def\url#1{\texttt{#1}}\fi
\expandafter\ifx\csname urlprefix\endcsname\relax\def\urlprefix{URL }\fi
\providecommand{\bibinfo}[2]{#2}
\providecommand{\eprint}[2][]{\url{#2}}
%
\bibitem[{\citenamefont{Kogut}(2007)}]{Kogut}
\bibinfo{author}{\bibfnamefont{J.}~\bibnamefont{B.}~\bibnamefont{Kogut}} 
\bibnamefont{and} 
\bibinfo{author}{\bibfnamefont{D.}~\bibnamefont{K.}~\bibnamefont{Sinclair}}  
\bibinfo{journal}{Phys. Rev.\  D} \textbf{\bibinfo{volume}{77}},
\bibinfo{pages}{114503} (\bibinfo{year}{2008}).
%
%
\bibitem[{\citenamefont{Fodor}(2002)}]{Fodor}
\bibinfo{author}{\bibfnamefont{Z.}~\bibnamefont{Fodor}}, 
\bibnamefont{and} 
\bibinfo{author}{\bibfnamefont{S.}~\bibnamefont{D.}~\bibnamefont{Katz}},  
  \bibinfo{journal}{Phys. Lett.\ B} \textbf{\bibinfo{volume}{534}},
  \bibinfo{pages}{87} (\bibinfo{year}{2002});
  \bibinfo{journal}{J. High Energy Phys.} \textbf{\bibinfo{volume}{03}},
  \bibinfo{pages}{014} (\bibinfo{year}{2002}).
%
\bibitem[{\citenamefont{Allton}(2004)}]{Allton}
\bibinfo{author}{\bibfnamefont{C.}~\bibfnamefont{R.}~\bibnamefont{Allton}},
\bibinfo{author}{\bibfnamefont{S.}~\bibnamefont{Ejiri}},
\bibinfo{author}{\bibfnamefont{S.}~\bibfnamefont{J.}~\bibnamefont{Hands}},
\bibinfo{author}{\bibfnamefont{O.}~\bibnamefont{Kaczmarek}},
\bibinfo{author}{\bibfnamefont{F.}~\bibnamefont{Karsch}},
\bibinfo{author}{\bibfnamefont{E.}~\bibnamefont{Laermann}},
\bibinfo{author}{\bibfnamefont{Ch.}~\bibnamefont{Schmidt}},
\bibnamefont{and} 
\bibinfo{author}{\bibfnamefont{L.}~\bibnamefont{Scorzato}},
  \bibinfo{journal}{Phys. Rev. D} \textbf{\bibinfo{volume}{66}},
  \bibinfo{pages}{074507} (\bibinfo{year}{2002});
\bibinfo{author}{\bibfnamefont{S.}~\bibnamefont{Ejiri}}, 
\bibinfo{author}{\bibfnamefont{C.}~\bibfnamefont{R.}~\bibnamefont{Allton}}, 
\bibinfo{author}{\bibfnamefont{S.}~\bibfnamefont{J.}~\bibnamefont{Hands}}, 
\bibinfo{author}{\bibfnamefont{O.}~\bibnamefont{Kaczmarek}},
\bibinfo{author}{\bibfnamefont{F.}~\bibnamefont{Karsch}},
\bibinfo{author}{\bibfnamefont{E.}~\bibnamefont{Laermann}},
\bibnamefont{and} 
\bibinfo{author}{\bibfnamefont{C.}~\bibnamefont{Schmidt}},  
  \bibinfo{journal}{Prog. Theor. Phys. Suppl.} \textbf{\bibinfo{volume}{153}},
  \bibinfo{pages}{118} (\bibinfo{year}{2004}).
%
\bibitem[{\citenamefont{Forcrand and Philipsen}(2002)}]{FP}
\bibinfo{author}{\bibfnamefont{P.}~\bibnamefont{de}~\bibnamefont{Forcrand}} 
\bibnamefont{and}
\bibinfo{author}{\bibfnamefont{O.}~\bibnamefont{Philipsen}},  
\bibinfo{journal}{Nucl. Phys. } \textbf{\bibinfo{volume}{B642}},
\bibinfo{pages}{290} (\bibinfo{year}{2002});
%
\bibitem[{\citenamefont{Forcrand and Philipsen}(2002)}]{FP3}
\bibinfo{author}{\bibfnamefont{P.}~\bibnamefont{de}~\bibnamefont{Forcrand}} 
\bibnamefont{and}
\bibinfo{author}{\bibfnamefont{O.}~\bibnamefont{Philipsen}},  
\bibinfo{journal}{Nucl. Phys. } \textbf{\bibinfo{volume}{B673}},
\bibinfo{pages}{170} (\bibinfo{year}{2003}). 
%
\bibitem[{\citenamefont{Elia and Lombardo}(2003)}]{Elia}
\bibinfo{author}{\bibfnamefont{M.}~\bibnamefont{D'Elia}} \bibnamefont{and}
\bibinfo{author}{\bibfnamefont{M.}~\bibfnamefont{P.}~\bibnamefont{Lombardo}},  
\bibinfo{journal}{Phys. Rev.\  D} \textbf{\bibinfo{volume}{67}},
\bibinfo{pages}{014505} (\bibinfo{year}{2003});
\bibinfo{journal}{Phys. Rev.\ D} \textbf{\bibinfo{volume}{70}},
\bibinfo{pages}{074509} (\bibinfo{year}{2004});
\bibinfo{author}{\bibfnamefont{M.}~\bibnamefont{D'Elia}},
 \bibinfo{author}{\bibfnamefont{F.}~\bibfnamefont{D.}~\bibnamefont{Renzo}},
\bibnamefont{and}
\bibinfo{author}{\bibfnamefont{M.}~\bibfnamefont{P.}~\bibnamefont{Lombardo}},  
\bibinfo{journal}{Phys. Rev.\ D} \textbf{\bibinfo{volume}{76}},
\bibinfo{pages}{114509} (\bibinfo{year}{2007});
%
\bibitem[{\citenamefont{Chen and Luo}(2005)}]{Chen34}
\bibinfo{author}{\bibfnamefont{H.}~\bibfnamefont{S.}~\bibnamefont{Chen}} \bibnamefont{and}
\bibinfo{author}{\bibfnamefont{X.}~\bibfnamefont{Q.}~\bibnamefont{Luo}},  
\bibinfo{journal}{Phys. Rev.} \textbf{\bibinfo{volume}{D72}},
\bibinfo{pages}{034504} (\bibinfo{year}{2005}); 
\bibinfo{howpublished}{arXiv:hep-lat/0702025} (\bibinfo{year}{2007}). 
%
\bibitem[{\citenamefont{Chen and Luo}(2005)}]{Chen}
\bibinfo{author}{\bibfnamefont{L.}~\bibfnamefont{K.}~\bibnamefont{Wu}}, 
\bibinfo{author}{\bibfnamefont{X.}~\bibfnamefont{Q.}~\bibnamefont{Luo}},  
\bibnamefont{and}
\bibinfo{author}{\bibfnamefont{H.}~\bibfnamefont{S.}~\bibnamefont{Chen}}, 
\bibinfo{journal}{Phys. Rev.} \textbf{\bibinfo{volume}{D76}},
\bibinfo{pages}{034505} (\bibinfo{year}{2007}). 
%
\bibitem[{\citenamefont{Nambu and Jona-Lasinio}(1961{\natexlab{a}})}]{NJ1}
\bibinfo{author}{\bibfnamefont{Y.}~\bibnamefont{Nambu}} \bibnamefont{and}
  \bibinfo{author}{\bibfnamefont{G.}~\bibnamefont{Jona-Lasinio}},
  \bibinfo{journal}{Phys.\ Rev.} \textbf{\bibinfo{volume}{122}},
  \bibinfo{pages}{345} (\bibinfo{year}{1961}); 
  \bibinfo{journal}{Phys.\ Rev.} \textbf{\bibinfo{volume}{124}},
  \bibinfo{pages}{246} (\bibinfo{year}{1961}).
%
\bibitem[{\citenamefont{Asakawa and Yazaki}(1989)}]{AY}
\bibinfo{author}{\bibfnamefont{M.}~\bibnamefont{Asakawa}} \bibnamefont{and}
  \bibinfo{author}{\bibfnamefont{K.}~\bibnamefont{Yazaki}},
  \bibinfo{journal}{Nucl.\ Phys.} \textbf{\bibinfo{volume}{A504}},
  \bibinfo{pages}{668} (\bibinfo{year}{1989}). 
%
\bibitem[{\citenamefont{Berges and Rajagopal}(1999)}]{BR}
\bibinfo{author}{\bibfnamefont{J.}~\bibnamefont{Berges}} \bibnamefont{and}
  \bibinfo{author}{\bibfnamefont{K.}~\bibnamefont{Rajagopal}},
  \bibinfo{journal}{Nucl.\ Phys.} \textbf{\bibinfo{volume}{B538}},
  \bibinfo{pages}{215} (\bibinfo{year}{1999}).
%
\bibitem[{\citenamefont{Scavenius et~al.}(2000)\citenamefont{Scavenius, {{\'
  A}. M{\' o}csy}, {I. N. Mishustin}, and {D. H. Rischke}}}]{Sca}
\bibinfo{author}{\bibfnamefont{O.}~\bibnamefont{Scavenius}},
  \bibinfo{author}{\bibnamefont{{{\' A}. M{\' o}csy}}},
  \bibinfo{author}{\bibnamefont{{I. N. Mishustin}}}, \bibnamefont{and}
  \bibinfo{author}{\bibnamefont{{D. H. Rischke}}}, \bibinfo{journal}{Phys.\
  Rev.\ C} \textbf{\bibinfo{volume}{64}}, \bibinfo{pages}{045202}
  (\bibinfo{year}{2001}).
%
\bibitem[{\citenamefont{Fujii}(2003)}]{Fuj}
\bibinfo{author}{\bibfnamefont{H.}~\bibnamefont{Fujii}},
\bibinfo{journal}{Phys.\ Rev.\ D} \textbf{\bibinfo{volume}{67}},
\bibinfo{pages}{094018} (\bibinfo{year}{2003}).
%
\bibitem[{\citenamefont{Kitazawa et al. }(2002)}]{KKKN}
\bibinfo{author}{\bibfnamefont{M.}~\bibnamefont{Kitazawa}},
{\bibfnamefont{T.}~\bibnamefont{Koide}},
{\bibfnamefont{T.}~\bibnamefont{Kunihiro}},
\bibnamefont{and}
\bibinfo{author}{\bibfnamefont{Y.}~\bibnamefont{Nemoto}},
\bibinfo{journal}{Prog. Theor. Phys.} \textbf{\bibinfo{volume}{108}},
\bibinfo{pages}{929} (\bibinfo{year}{2002}). 
%
\bibitem[{\citenamefont{Osipov et al. }(2006)}]{Osipov}
\bibinfo{author}{\bibfnamefont{A.}~\bibfnamefont{A.}~\bibnamefont{Osipov}},
{\bibfnamefont{B.}~\bibnamefont{Hiller}},
\bibnamefont{and} 
{\bibfnamefont{J.}~\bibnamefont{da Provid\^encia}},
\bibinfo{journal}{Phys. Lett.\ B} \textbf{\bibinfo{volume}{634}},
\bibinfo{pages}{48} (\bibinfo{year}{2006});
\bibinfo{author}{\bibfnamefont{A.}~\bibfnamefont{A.}~\bibnamefont{Osipov}},
{\bibfnamefont{B.}~\bibnamefont{Hiller}}, 
{\bibfnamefont{J.}~\bibnamefont{Moreira}}, 
\bibnamefont{and} 
{\bibfnamefont{A.}~\bibfnamefont{H.}~\bibnamefont{Blin}},
\bibinfo{journal}{Eur. Phys. J.\ C} \textbf{\bibinfo{volume}{46}},
\bibinfo{pages}{225} (\bibinfo{year}{2006}); 
\bibinfo{author}{\bibfnamefont{A.}~\bibfnamefont{A.}~\bibnamefont{Osipov}},
{\bibfnamefont{B.}~\bibnamefont{Hiller}},
{\bibfnamefont{J.}~\bibnamefont{Moreira}},
{\bibfnamefont{A.}~\bibnamefont{H.}~\bibnamefont~{Blin}},
\bibnamefont{and} 
{\bibfnamefont{J.}~\bibnamefont{da Provid\^encia}},
\bibinfo{journal}{Phys. Lett.\ B} \textbf{\bibinfo{volume}{646}},
\bibinfo{pages}{91} (\bibinfo{year}{2007}); 
\bibinfo{author}{\bibfnamefont{A.}~\bibfnamefont{A.}~\bibnamefont{Osipov}},
\bibinfo{author}{\bibfnamefont{B.}~\bibnamefont{Hiller}},
\bibinfo{author}{\bibfnamefont{J.}~\bibnamefont{Moreira}},
\bibnamefont{and}
\bibinfo{author}{\bibfnamefont{A.}~\bibfnamefont{H.}~\bibnamefont{Blin}}, 
\bibinfo{journal}{Phys. Lett.\ B} \textbf{\bibinfo{volume}{659}},
\bibinfo{pages}{270} (\bibinfo{year}{2008});
\bibinfo{author}{\bibfnamefont{B.}~\bibnamefont{Hiller}},
\bibinfo{author}{\bibfnamefont{A.}~\bibfnamefont{A.}~\bibnamefont{Osipov}},
\bibinfo{author}{\bibfnamefont{A.}~\bibfnamefont{H.}~\bibnamefont{Blin}},
\bibnamefont{and} 
{\bibfnamefont{J.}~\bibnamefont{da Provid\^encia}},
\bibinfo{howpublished}{arXiv:hep-ph/0802.3193} (\bibinfo{year}{2008});
\bibinfo{author}{\bibfnamefont{B.}~\bibnamefont{Hiller}},
\bibinfo{author}{\bibfnamefont{A.}~\bibfnamefont{A.}~\bibnamefont{Osipov}},
\bibinfo{author}{\bibfnamefont{J.}~\bibnamefont{Moreira}},
\bibnamefont{and} 
\bibinfo{author}{\bibfnamefont{A.}~\bibfnamefont{H.}~\bibnamefont{Blin}},
\bibinfo{howpublished}{arXiv:hep-ph/0809.2515} (\bibinfo{year}{2008});
\bibinfo{author}{\bibfnamefont{B.}~\bibnamefont{Hiller}},
\bibinfo{author}{\bibfnamefont{J.}~\bibnamefont{Moreira}},
\bibinfo{author}{\bibfnamefont{A.}~\bibfnamefont{A.}~\bibnamefont{Osipov}},
\bibnamefont{and} 
\bibinfo{author}{\bibfnamefont{A.}~\bibfnamefont{H.}~\bibnamefont{Blin}},
\bibinfo{howpublished}{arXiv:hep-ph/0812.1532} (\bibinfo{year}{2008}). 
%
\bibitem[{\citenamefont{Kashiwa et al}(2006)}]{Kashiwa}
\bibinfo{author}{\bibfnamefont{K.}~\bibnamefont{Kashiwa}}, 
\bibinfo{author}{\bibfnamefont{H.}~\bibnamefont{Kouno}}, 
\bibinfo{author}{\bibfnamefont{T.}~\bibnamefont{Sakaguchi}}, 
\bibinfo{author}{\bibfnamefont{M.}~\bibnamefont{Matsuzaki}}, 
\bibnamefont{and}
\bibinfo{author}{\bibfnamefont{M.}~\bibnamefont{Yahiro}},
\bibinfo{journal}{Phys. Lett.\ B} \textbf{\bibinfo{volume}{647}},
\bibinfo{pages}{446} (\bibinfo{year}{2007}); 
\bibinfo{author}{\bibfnamefont{K.}~\bibnamefont{Kashiwa}}, 
\bibinfo{author}{\bibfnamefont{M.}~\bibnamefont{Matsuzaki}}, 
\bibinfo{author}{\bibfnamefont{H.}~\bibnamefont{Kouno}}, 
\bibnamefont{and}
\bibinfo{author}{\bibfnamefont{M.}~\bibnamefont{Yahiro}},
\bibinfo{journal}{Phys. Lett.\ B} \textbf{\bibinfo{volume}{657}},
\bibinfo{pages}{143} (\bibinfo{year}{2007}). 
%
\bibitem[{\citenamefont{Sakaguchi et al}(2006)}]{Sakaetal}
\bibinfo{author}{\bibfnamefont{T.}~\bibnamefont{Sakaguchi}}, 
\bibinfo{author}{\bibfnamefont{M.}~\bibnamefont{Matsuzaki}}, 
\bibinfo{author}{\bibfnamefont{H.}~\bibnamefont{Kouno}}, 
 \bibnamefont{and}
\bibinfo{author}{\bibfnamefont{M.}~\bibnamefont{Yahiro}},
\bibinfo{journal}{Centr.\ Eur.\ J.\ Phys.} \textbf{\bibinfo{volume}{6}},
\bibinfo{pages}{116} (\bibinfo{year}{2008}).
%
\bibitem[{\citenamefont{Meisinger et al.}(1996)}]{Meisinger}
\bibinfo{author}{\bibfnamefont{P.}~\bibnamefont{N.}}~\bibnamefont{Meisinger},
\bibnamefont{and}
\bibinfo{author}{\bibfnamefont{M.}~\bibnamefont{C.}}~\bibnamefont{Ogilvie},  
  \bibinfo{journal}{Phys. Lett.\ B} \textbf{\bibinfo{volume}{379}},
  \bibinfo{pages}{163} (\bibinfo{year}{1996}). 
%
\bibitem[{\citenamefont{Dumitru}(2002)}]{Dumitru}
\bibinfo{author}{\bibfnamefont{A.}~\bibnamefont{Dumitru}},
\bibnamefont{and}
\bibinfo{author}{\bibfnamefont{R.}~\bibfnamefont{D.}~\bibnamefont{Pisarski}},  
\bibinfo{journal}{Phys.\ Rev.\  D} \textbf{\bibinfo{volume}{66}},
\bibinfo{pages}{096003} (\bibinfo{year}{2002}); 
\bibinfo{author}{\bibfnamefont{A.}~\bibnamefont{Dumitru}},
\bibinfo{author}{\bibfnamefont{Y.}~\bibnamefont{Hatta}},
\bibinfo{author}{\bibfnamefont{J.}~\bibnamefont{Lenaghan}},
\bibinfo{author}{\bibfnamefont{K.}~\bibnamefont{Orginos}},
\bibnamefont{and}
\bibinfo{author}{\bibfnamefont{R.}~\bibfnamefont{D.}~\bibnamefont{Pisarski}},  
\bibinfo{journal}{Phys.\ Rev.\  D} \textbf{\bibinfo{volume}{70}},
\bibinfo{pages}{034511} (\bibinfo{year}{2004}); 
\bibinfo{author}{\bibfnamefont{A.}~\bibnamefont{Dumitru}},
\bibinfo{author}{\bibfnamefont{R.}~\bibfnamefont{D.}~\bibnamefont{Pisarski}},  
\bibnamefont{and}
\bibinfo{author}{\bibfnamefont{D.}~\bibnamefont{Zschiesche}},  
\bibinfo{journal}{Phys.\ Rev.\  D} \textbf{\bibinfo{volume}{72}},
\bibinfo{pages}{065008} (\bibinfo{year}{2005}).
%
\bibitem[{\citenamefont{Fukushima}(2004)}]{Fukushima}
\bibinfo{author}{\bibfnamefont{K.}~\bibnamefont{Fukushima}}, 
  \bibinfo{journal}{Phys. Lett.\ B} \textbf{\bibinfo{volume}{591}},
  \bibinfo{pages}{277} (\bibinfo{year}{2004}).
%
\bibitem[{\citenamefont{Fukushima}(2008)}]{Fukushima2}
\bibinfo{author}{\bibfnamefont{K.}~\bibnamefont{Fukushima}}, 
  \bibinfo{journal}{Phys.\ Rev.\  D} \textbf{\bibinfo{volume}{77}},
  \bibinfo{pages}{114028} (\bibinfo{year}{2008}).
%
\bibitem[{\citenamefont{{S. K. Ghosh} et al.}(2006)}]{Ghos}
\bibinfo{author}{\bibnamefont{{S. K. Ghosh}}},
  \bibinfo{author}{\bibnamefont{{T. K. Mukherjee}}},
  \bibinfo{author}{\bibnamefont{{M. G. Mustafa}}}, \bibnamefont{and}
  \bibinfo{author}{\bibfnamefont{R.}~\bibnamefont{Ray}},
  \bibinfo{journal}{Phys.\ Rev.\ D} \textbf{\bibinfo{volume}{73}},
  \bibinfo{pages}{114007} (\bibinfo{year}{2006}). 
%
\bibitem[{\citenamefont{Megias et al.}(2006)}]{Megias}
\bibinfo{author}{\bibfnamefont{E.}~\bibnamefont{Meg{$\acute{\i}$}as}},
\bibinfo{author}{\bibfnamefont{E.}~\bibnamefont{R.}~\bibnamefont{Arriola}},
\bibnamefont{and}
\bibinfo{author}{\bibfnamefont{L.}~\bibnamefont{L.}~\bibnamefont{Salcedo}},  
  \bibinfo{journal}{Phys. Rev.\ D} \textbf{\bibinfo{volume}{74}},
  \bibinfo{pages}{065005} (\bibinfo{year}{2006}). 
%
\bibitem[{\citenamefont{Ratti et al.}(2006)}]{Ratti}
\bibinfo{author}{\bibfnamefont{C.}~\bibnamefont{Ratti}},
\bibinfo{author}{\bibfnamefont{M.}~\bibfnamefont{A.}~\bibnamefont{Thaler}},
\bibnamefont{and}
\bibinfo{author}{\bibfnamefont{W.}~\bibnamefont{Weise}},  
  \bibinfo{journal}{Phys. Rev.\ D} \textbf{\bibinfo{volume}{73}},
  \bibinfo{pages}{014019} (\bibinfo{year}{2006}); 
\bibinfo{author}{\bibfnamefont{C.}~\bibnamefont{Ratti}},
\bibinfo{author}{\bibfnamefont{S.}~\bibnamefont{R\"{o}{\ss}ner}},
\bibinfo{author}{\bibfnamefont{M.}~\bibfnamefont{A.}~\bibnamefont{Thaler}},
\bibnamefont{and}
\bibinfo{author}{\bibfnamefont{W.}~\bibnamefont{Weise}},  
  \bibinfo{journal}{Eur. Phys. J.\ C} \textbf{\bibinfo{volume}{49}},
  \bibinfo{pages}{213} (\bibinfo{year}{2007}). 
%
\bibitem[{\citenamefont{Rossner et al.}(2007)}]{Rossner}
\bibinfo{author}{\bibfnamefont{S.}~\bibnamefont{R\"{o}{\ss}ner}},
\bibinfo{author}{\bibfnamefont{C.}~\bibnamefont{Ratti}},
\bibnamefont{and}
\bibinfo{author}{\bibfnamefont{W.}~\bibnamefont{Weise}},  
  \bibinfo{journal}{Phys. Rev.\ D} \textbf{\bibinfo{volume}{75}},
  \bibinfo{pages}{034007} (\bibinfo{year}{2007}). 
%
\bibitem[{\citenamefont{Ciminale}(2007)}]{Ciminale}
\bibinfo{author}{\bibfnamefont{M.}~\bibnamefont{Ciminale}},
\bibinfo{author}{\bibfnamefont{R.}~\bibnamefont{Gatto}},
\bibinfo{author}{\bibfnamefont{N.}~\bibfnamefont{D.}~\bibnamefont{Ippolito}},
\bibinfo{author}{\bibfnamefont{G.}~\bibnamefont{Nardulli}},  
\bibnamefont{and}
\bibinfo{author}{\bibfnamefont{M.}~\bibnamefont{Ruggieri}},  
  \bibinfo{journal}{Phys. Rev.\ D} \textbf{\bibinfo{volume}{77}},
  \bibinfo{pages}{054023} (\bibinfo{year}{2008});
\bibinfo{author}{\bibfnamefont{M.}~\bibnamefont{Ciminale}},
\bibinfo{author}{\bibfnamefont{G.}~\bibnamefont{Nardulli}},  
\bibinfo{author}{\bibfnamefont{M.}~\bibnamefont{Ruggieri}},
\bibnamefont{and}
\bibinfo{author}{\bibfnamefont{R.}~\bibnamefont{Gatto}},
  \bibinfo{journal}{Phys.\ Lett.\ B} \textbf{\bibinfo{volume}{657}},
  \bibinfo{pages}{64} (\bibinfo{year}{2007}).
%
\bibitem[{\citenamefont{Hansen et al.}(2007)}]{Hansen}
\bibinfo{author}{\bibfnamefont{H.}~\bibnamefont{Hansen}}, 
\bibinfo{author}{\bibfnamefont{W.}~\bibfnamefont{M.}~\bibnamefont{Alberico}},
\bibinfo{author}{\bibfnamefont{A.}~\bibnamefont{Beraudo}}, 
\bibinfo{author}{\bibfnamefont{A.}~\bibnamefont{Molinari}},
\bibinfo{author}{\bibfnamefont{M.}~\bibnamefont{Nardi}},
\bibnamefont{and}
\bibinfo{author}{\bibfnamefont{C.}~\bibnamefont{Ratti}}, 
  \bibinfo{journal}{Phys. Rev.\ D} \textbf{\bibinfo{volume}{75}},
  \bibinfo{pages}{065004} (\bibinfo{year}{2007}). 
%
\bibitem[{\citenamefont{Sasaki et al.}(2007)}]{Sasaki}
\bibinfo{author}{\bibfnamefont{C.}~\bibnamefont{Sasaki}},
\bibinfo{author}{\bibfnamefont{B.}~\bibnamefont{Friman}},
\bibnamefont{and}
\bibinfo{author}{\bibfnamefont{K.}~\bibnamefont{Redlich}}, 
\bibinfo{journal}{Phys. Rev.\ D} \textbf{\bibinfo{volume}{75}},
  \bibinfo{pages}{074013} (\bibinfo{year}{2007}). 
%
\bibitem[{\citenamefont{Schaefer}(2007)}]{Schaefer}
\bibinfo{author}{\bibfnamefont{B.}~\bibfnamefont{-J.}~\bibnamefont{Schaefer}},
\bibinfo{author}{\bibfnamefont{J.}~\bibfnamefont{M.}~\bibnamefont{Pawlowski}},
\bibnamefont{and}
\bibinfo{author}{\bibfnamefont{J.}~\bibnamefont{Wambach}},  
  \bibinfo{journal}{Phys.\ Rev.\  D} \textbf{\bibinfo{volume}{76}},
  \bibinfo{pages}{074023} (\bibinfo{year}{2007}).

%
\bibitem[{\citenamefont{Costa et al}(2008)}]{Costa}
\bibinfo{author}{\bibfnamefont{P.}~\bibnamefont{Costa}}, 
\bibinfo{author}{\bibfnamefont{C.}~\bibfnamefont{A.}~\bibfnamefont{de}~\bibnamefont{Sousa}}, 
\bibinfo{author}{\bibfnamefont{M.}~\bibfnamefont{C.}~\bibnamefont{Ruivo}}, 
\bibnamefont{and}
\bibinfo{author}{\bibfnamefont{H.}~\bibnamefont{Hansen}},
 \bibinfo{howpublished}{arXiv:hep-ph/0801.3616} (\bibinfo{year}{2008});
\bibinfo{author}{\bibfnamefont{P.}~\bibnamefont{Costa}}, 
\bibinfo{author}{\bibfnamefont{M.}~\bibfnamefont{C.}~\bibnamefont{Ruivo}}, 
\bibinfo{author}{\bibfnamefont{C.}~\bibfnamefont{A.}~\bibfnamefont{de}~\bibnamefont{Sousa}}, 
\bibinfo{author}{\bibfnamefont{H.}~\bibnamefont{Hansen}},
\bibnamefont{and}
\bibinfo{author}{\bibfnamefont{W.}~\bibfnamefont{M.}~\bibnamefont{Alberico}},
 \bibinfo{howpublished}{arXiv:hep-ph/0807.2134} (\bibinfo{year}{2008}). 
%
\bibitem[{\citenamefont{Kashiwa et al}(2008)}]{Kashiwa1}
\bibinfo{author}{\bibfnamefont{K.}~\bibnamefont{Kashiwa}}, 
\bibinfo{author}{\bibfnamefont{H.}~\bibnamefont{Kouno}}, 
\bibinfo{author}{\bibfnamefont{M.}~\bibnamefont{Matsuzaki}}, 
\bibnamefont{and}
\bibinfo{author}{\bibfnamefont{M.}~\bibnamefont{Yahiro}},
  \bibinfo{journal}{Phys.\ Lett.\ B} \textbf{\bibinfo{volume}{662}},
  \bibinfo{pages}{26} (\bibinfo{year}{2008}).
%
\bibitem[{\citenamefont{Fu}(2007)}]{Fu}
\bibinfo{author}{\bibfnamefont{W.}~\bibfnamefont{J.}~\bibnamefont{Fu}},
\bibinfo{author}{\bibfnamefont{Z.}~\bibnamefont{Zhang}},
\bibnamefont{and}
\bibinfo{author}{\bibfnamefont{Y.}~\bibfnamefont{X.}~\bibnamefont{Liu}},
  \bibinfo{journal}{Phys.\ Rev.\  D} \textbf{\bibinfo{volume}{77}},
  \bibinfo{pages}{014006} (\bibinfo{year}{2008}).
%
\bibitem[{\citenamefont{Abuki}(2008)}]{Abuki}
\bibinfo{author}{\bibfnamefont{H.}~\bibnamefont{Abuki}},
\bibinfo{author}{\bibfnamefont{M.}~\bibnamefont{Ciminale}},
\bibinfo{author}{\bibfnamefont{R.}~\bibnamefont{Gatto}},
\bibinfo{author}{\bibfnamefont{G.}~\bibnamefont{Nardulli}},
\bibnamefont{and}
\bibinfo{author}{\bibfnamefont{M.}~\bibnamefont{Ruggieri}},
 \bibinfo{journal}{Phys.\ Rev.\  D} \textbf{\bibinfo{volume}{77}},
  \bibinfo{pages}{074018} (\bibinfo{year}{2008});
\bibinfo{author}{\bibfnamefont{H.}~\bibnamefont{Abuki}},
\bibinfo{author}{\bibfnamefont{M.}~\bibnamefont{Ciminale}},
\bibinfo{author}{\bibfnamefont{R.}~\bibnamefont{Gatto}},
\bibinfo{author}{\bibfnamefont{N.}~\bibfnamefont{D.}~\bibnamefont{Ippolito}},
\bibinfo{author}{\bibfnamefont{G.}~\bibnamefont{Nardulli}},
\bibnamefont{and}
\bibinfo{author}{\bibfnamefont{M.}~\bibnamefont{Ruggieri}},
 \bibinfo{journal}{Phys.\ Rev.\  D} \textbf{\bibinfo{volume}{78}},
  \bibinfo{pages}{014002} (\bibinfo{year}{2008});  
\bibinfo{author}{\bibfnamefont{H.}~\bibnamefont{Abuki}},
\bibinfo{author}{\bibfnamefont{R.}~\bibnamefont{Anglani}},
\bibinfo{author}{\bibfnamefont{R.}~\bibnamefont{Gatto}},
\bibinfo{author}{\bibfnamefont{G.}~\bibnamefont{Nardulli}},
\bibnamefont{and}
\bibinfo{author}{\bibfnamefont{G.}~\bibnamefont{Nardulli}},
 \bibinfo{journal}{Phys.\ Rev.\  D} \textbf{\bibinfo{volume}{78}},
  \bibinfo{pages}{034034} (\bibinfo{year}{2008}).   
%
\bibitem[{\citenamefont{Sakai et al}(2008)}]{Sakai}
\bibinfo{author}{\bibfnamefont{Y.}~\bibnamefont{Sakai}},
\bibinfo{author}{\bibfnamefont{K.}~\bibnamefont{Kashiwa}}, 
\bibinfo{author}{\bibfnamefont{H.}~\bibnamefont{Kouno}}, 
\bibnamefont{and}
\bibinfo{author}{\bibfnamefont{M.}~\bibnamefont{Yahiro}},
  \bibinfo{journal}{Phys.\ Rev.\  D} \textbf{\bibinfo{volume}{77}},
  \bibinfo{pages}{051901(R)} (\bibinfo{year}{2008});
  \bibinfo{journal}{Phys.\ Rev.\  D} \textbf{\bibinfo{volume}{78}},
  \bibinfo{pages}{ 036001} (\bibinfo{year}{2008}). 
%
\bibitem[{\citenamefont{Sakai et al}(2008)}]{Sakai1}
\bibinfo{author}{\bibfnamefont{Y.}~\bibnamefont{Sakai}},
\bibinfo{author}{\bibfnamefont{K.}~\bibnamefont{Kashiwa}}, 
\bibinfo{author}{\bibfnamefont{H.}~\bibnamefont{Kouno}}, 
\bibinfo{author}{\bibfnamefont{M.}~\bibnamefont{Matsuzaki}}, 
\bibnamefont{and}
\bibinfo{author}{\bibfnamefont{M.}~\bibnamefont{Yahiro}},  
  \bibinfo{journal}{Phys.\ Rev.\  D} \textbf{\bibinfo{volume}{78}},
  \bibinfo{pages}{ 076007} (\bibinfo{year}{2008}).
%
\bibitem[{\citenamefont{Roberge and Weiss}(1986)}]{RW}
\bibinfo{author}{\bibfnamefont{A.}~\bibnamefont{Roberge}} \bibnamefont{and}
\bibinfo{author}{\bibfnamefont{N.}~\bibnamefont{Weiss}},  
\bibinfo{journal}{Nucl. Phys. } \textbf{\bibinfo{volume}{B275}},
\bibinfo{pages}{734} (\bibinfo{year}{1986}). 
%
\bibitem[{\citenamefont{Walecka}(1974)}]{Walecka}
\bibinfo{author}{\bibfnamefont{J. D.}~\bibnamefont{Walecka}},
\bibinfo{journal}{Ann. Phys. } \textbf{\bibinfo{volume}{83}},
\bibinfo{pages}{491} (\bibinfo{year}{1974}).
%
\bibitem[{\citenamefont{Kashiwa and Sakaguchi}(2003)}]{KS}
\bibinfo{author}{\bibfnamefont{T.}~\bibnamefont{Kashiwa}} \bibnamefont{and}
\bibinfo{author}{\bibfnamefont{T.}~\bibnamefont{Sakaguchi}},
\bibinfo{journal}{Phys.\ Rev.\ D} \textbf{\bibinfo{volume}{68}},
\bibinfo{pages}{065002} (\bibinfo{year}{2003}).
%
\bibitem[{\citenamefont{Kouno et al.}(2005)}]{KSKHTMY}
\bibinfo{author}{\bibfnamefont{H.}~\bibnamefont{Kouno}},
\bibinfo{author}{\bibfnamefont{T.}~\bibnamefont{Sakaguchi}},
\bibinfo{author}{\bibfnamefont{K.}~\bibnamefont{Kashiwa}},
\bibinfo{author}{\bibfnamefont{M.}~\bibnamefont{Hamada}},
\bibinfo{author}{\bibfnamefont{H.}~\bibnamefont{Tokudome}},
\bibinfo{author}{\bibfnamefont{M.}~\bibnamefont{Matsuzaki}},
\bibnamefont{and}
\bibinfo{author}{\bibfnamefont{M.}~\bibnamefont{Yahiro}},
\bibinfo{journal}{Soryushiron Kenkyu} \textbf{\bibinfo{volume}{112}}
\bibinfo{pages}{C67}
\bibinfo{howpublished}{(arXiv:nucl-th/0509057)} 
 (\bibinfo{year}{2005}).
%
\bibitem[{\citenamefont{Karsch}(2002)}]{Karsch3}
\bibinfo{author}{\bibfnamefont{F.}~\bibnamefont{Karsch}}, 
\bibinfo{journal}{Lect. notes Phys. } \textbf{\bibinfo{volume}{583}},
\bibinfo{pages}{209} (\bibinfo{year}{2002}). 
%
\bibitem[{\citenamefont{Karsch, Leermann and Peikert}(2002)}]{Karsch4}
\bibinfo{author}{\bibfnamefont{F.}~\bibnamefont{Karsch}}, 
\bibinfo{author}{\bibfnamefont{E.}~\bibnamefont{Laermann}}, 
\bibnamefont{and}
\bibinfo{author}{\bibfnamefont{A.}~\bibnamefont{Peikert}},
\bibinfo{journal}{Nucl. Phys. \ B} \textbf{\bibinfo{volume}{605}},
\bibinfo{pages}{579} (\bibinfo{year}{2002}). 
%
\bibitem{Kaczmarek2}
\bibinfo{author}{\bibfnamefont{M.}~\bibnamefont{Kaczmarek}}
\bibnamefont{and}
\bibinfo{author}{\bibfnamefont{F.}~\bibnamefont{Zantow}}, 
  \bibinfo{journal}{Phys.\ Rev.\  D} \textbf{\bibinfo{volume}{71}},
  \bibinfo{pages}{114510} (\bibinfo{year}{2005});
%
\bibitem[{\citenamefont{Boyd et al.}(1996)}]{Boyd}
\bibinfo{author}{\bibfnamefont{G.}~\bibnamefont{Boyd}},
\bibinfo{author}{\bibfnamefont{J.}~\bibnamefont{Engels}},
\bibinfo{author}{\bibfnamefont{F.}~\bibnamefont{Karsch}},
\bibinfo{author}{\bibfnamefont{E.}~\bibnamefont{Laermann}},
\bibinfo{author}{\bibfnamefont{C.}~\bibnamefont{Legeland}},
\bibinfo{author}{\bibfnamefont{M.}~\bibnamefont{L\"{u}tgemeier}},
\bibnamefont{and}
\bibinfo{author}{\bibfnamefont{B.}~\bibnamefont{Petersson}},
 \bibinfo{journal}{Nucl. Phys.} \textbf{\bibinfo{volume}{B469}},
\bibinfo{pages}{419} (\bibinfo{year}{1996}). 
%
\bibitem[{\citenamefont{Kaczmarek}(2002)}]{Kaczmarek}
\bibinfo{author}{\bibfnamefont{O.}~\bibnamefont{Kaczmarek}},
\bibinfo{author}{\bibfnamefont{F.}~\bibnamefont{Karsch}},
\bibinfo{author}{\bibfnamefont{P.}~\bibnamefont{Petreczky}},
\bibnamefont{and}
\bibinfo{author}{\bibfnamefont{F.}~\bibnamefont{Zantow}},  
  \bibinfo{journal}{Phys. Lett.\ B} \textbf{\bibinfo{volume}{543}},
  \bibinfo{pages}{41} (\bibinfo{year}{2002}).

\bibitem[{\citenamefont{Ejiri}(2007)}]{Ejiri}
\bibinfo{author}{\bibfnamefont{S.}~\bibnamefont{Ejiri}}, 
\bibinfo{howpublished}{arXiv:hep-lat/0706.3549} (\bibinfo{year}{2007}); 
\bibinfo{howpublished}{arXiv:hep-lat/0710.0653} (\bibinfo{year}{2007}).

\end{thebibliography}
\end{document}